\documentclass{llncs}
\usepackage{graphicx}
\DeclareGraphicsExtensions{.pdf,.png,.jpg}
\usepackage{amsfonts}
\usepackage[fleqn]{amsmath}
\usepackage{amssymb}
\usepackage{mathscinet}
\usepackage{tikz}
\usetikzlibrary{decorations.markings} 
\usepackage[T1]{fontenc}
\usepackage{scalefnt}
\usepackage{makeidx}
\usepackage{algorithm}
\usepackage{xspace}
\usepackage{wasysym}
\usepackage{stmaryrd}
\usepackage{calc}
\usepackage{hyperref} 

\usetikzlibrary{petri}
\usetikzlibrary{positioning}
\usetikzlibrary{backgrounds}
\tikzset{
LTS/.style={node distance=3em, on grid},
state/.style={circle,fill=black,inner sep=0.05cm},
edge/.style={-latex, auto},
}

\usepackage{filecontents} 
\tikzset{
    pos plot label/.style 2 args={
        postaction={
            decorate,
            decoration={
                markings,
                mark=at position #1 with \node#2;
            }
        }
    }
}
\makeatletter
\tikzset{
    scale plot marks/.is choice,
    scale plot marks/false/.code={
        \def\pgfuseplotmark##1{\pgftransformresetnontranslations\csname pgf@plot@mark@##1\endcsname}
    },
    scale plot marks/true/.style={},
    scale plot marks/.default=true
}
\makeatother

\tikzset{
    LTSArc/.style={-Latex},
    PNFlow/.style={-Triangle}
}

\makeindex

\begin{document}

\newcommand{\lts}{\ensuremath{\mathit{lts}}\xspace}
\newcommand{\PN}{\mathit{PN}}
\newcommand{\es}{\emptyset}
\DeclareRobustCommand{\pminus}{\mathbin{\ooalign{\hfil $-$\hfil \cr \hfil\raisebox{1.5mm}{$\scriptstyle\bullet$}\hfil}}} 
\newcommand{\leer}{\varepsilon}
\newcommand{\N}{\mathbb{N}}
\newcommand{\Z}{\mathbb{Z}}
\newcommand{\Q}{\mathbb{Q}}
\newcommand{\minus}{\setminus}
\newcommand{\emptyseq}{\varepsilon}
\newcommand{\then}{\Rightarrow}
\newcommand{\parikh}{{\mathcal P}}
\def\tp{^{\sf T}}
\newcommand{\tile}[4]{\begin{array}{|ll|}\hline#1&#2\\#3&#4\\\hline\end{array}}
\newcommand{\support}{\mathit{supp}}
\newcommand{\one}{\mathbf{1}}
\newcommand{\zero}{\ensuremath{\mathbf{0}}}
\newcommand{\step}[1]{[#1\rangle}
\newcommand{\M}{{\mathcal M}}
\newcommand{\comment}[1]{\hfill $\triangleright$ #1}
\newcounter{mylinenr}[algorithm]
\newlength{\myindent}
\newcommand{\linenr}[1][]{\stepcounter{mylinenr}\addtolength{\myindent}{#1em}\makebox[1em][r]{\scriptsize\arabic{mylinenr}:}\hspace*{\myindent}}
\newcommand{\NP}{{\sf NP}}
\newcommand{\PP}{{\sf P}}
\newcommand{\EX}{{\sf EXPSPACE}}

\newcommand{\disjcup}{\makebox[1.0em][c]{
\setlength{\unitlength}{0.4em}
\begin{picture}(2,2)(0,0)
\put(0.84,0.48){{\tiny $\bullet$}}
\put(0.5,0){$\cup$}
\end{picture}
\setlength{\unitlength}{1mm}
}}

\newcommand{\EndSy}{\hfill\protect\makebox[1.0em][c]{
\protect\setlength{\unitlength}{0.2em}
\protect\begin{picture}(3,3)(0,0)
        \begin{thinlines}
\protect\put(0,0){\line(1,0){3}}
\protect\put(0,0){\line(0,1){3}}
\protect\put(0,3){\line(1,0){3}}
\protect\put(3,0){\line(0,1){3}}
        \end{thinlines}
\protect\end{picture}
\protect\setlength{\unitlength}{1mm}
}}

\newcommand{\choice}{\protect\makebox[1.0em][c]{
\protect\setlength{\unitlength}{0.2em}
\protect\begin{picture}(2,4)(0,0)
\protect\put(0,0){\line(1,0){2}}
\protect\put(0,0){\line(0,0){4}}
\protect\put(0,4){\line(1,0){2}}
\protect\put(2,0){\line(0,1){4}}
\protect\end{picture}
\protect\setlength{\unitlength}{1mm}
}}

\newcommand{\BX}[1]{{\unskip\nobreak\hfil\penalty50
                    \hskip2em\hbox{}\hfil
\EndSy\/ {{\rm #1}}
                    \parfillskip=0pt \finalhyphendemerits=0 \par
                   }}

\newcommand{\DEF}[2]{\goodbreak\begin{definition}
                     \label{#1}\begin{rm}{\sc #2}

                    }
\newcommand{\ENDDEF}[1]{\BX{\ref{#1}}
                        \end{rm}\end{definition}
                       }
\newcommand{\ENXDEF}{\end{rm}\end{definition}}
\newcommand{\KRYPT}[2]{\goodbreak\begin{kryptosystem}
                     \label{#1}\begin{rm}{\sc #2}

                    }
\newcommand{\ENDKRYPT}[1]{\BX{\ref{#1}}
                        \end{rm}\end{kryptosystem}
                       }
\newcommand{\KOR}[2]{\goodbreak\begin{corollary}
                     \label{#1}{\sc #2}

                    }
\newcommand{\ENDKOR}[1]{\BX{\ref{#1}}
\end{corollary}
                       }
\newcommand{\ENXKOR}{
\end{corollary}}
\newtheorem{notation}[theorem]{{\bf Notation}} 
\newcommand{\PROP}[2]{\goodbreak\begin{proposition} 
                      \label{#1}{\sc #2}

                     }
\newcommand{\ENDPROP}{
\end{proposition}} 
\newcommand{\ENXPROP}[1]{\BX{\ref{#1}}
\end{proposition}} 
\newcommand{\THEO}[2]{\goodbreak\begin{theorem}
                     \label{#1}{\sc #2}

                    }
\newcommand{\ENDTHEO}{
\end{theorem}}
\newcommand{\SATZ}[2]{\goodbreak\begin{theorem}
                     \label{#1}{\sc #2}

                    }
\newcommand{\ENDSATZ}{
\end{theorem}}
\newcommand{\ENXSATZ}[1]{\BX{\ref{#1}}
\end{theorem}}
\newcommand{\LEM}[2]{\goodbreak\begin{lemma}
                     \label{#1}{\sc #2}

                    }
\newcommand{\ENDLEM}{
\end{lemma}}
\newcommand{\ENXLEM}[1]{\BX{\ref{#1}}
\end{lemma}}
\newcommand{\BEW}{\goodbreak
{\bf Proof:}
                 }
\newcommand{\XBEW}{\goodbreak{\bf Proof:} }
\newcommand{\ENDBEW}[1]{\BX{\ref{#1}}
                       }
\newcommand{\ENXBEW}{}
\newcommand{\BBEW}[1]{\goodbreak{\bf Beweis von {\rm #1}:}}

\newcommand{\ENDBBEW}[1]{\BX{\ref{#1}}
                        }
\newcommand{\ENXBBEW}{}
\newcommand{\NOT}[2]{\goodbreak\begin{notation}
                     \label{#1}\begin{rm}{\sc #2}

                    }
\newcommand{\ENDNOT}[1]{\BX{\ref{#1}}
                        \end{rm}\end{notation}
                       }
\newcommand{\ENXNOT}{\end{rm}\end{notation}}
\newcommand{\REM}[2]{\goodbreak\begin{remark}
                     \label{#1}\begin{rm}{\sc #2}
                    }
\newcommand{\ENDREM}[1]{\BX{\ref{#1}}
                        \end{rm}\end{remark}
                       }
\newcommand{\ENXREM}{\end{rm}\end{remark}}
\newcommand{\CON}[2]{\goodbreak\begin{conjecture}
                     \label{#1}\begin{rm}{\sc #2}
                    }
\newcommand{\ENDCON}[1]{\BX{\ref{#1}}
                        \end{rm}\end{conjecture}
                       }
\newcommand{\ENXCOM}{\end{rm}\end{conjecture}}
\newcommand{\BSP}[2]{\goodbreak\begin{beispiel}
                     \label{#1}\begin{rm}{\sc #2}

                    }
\newcommand{\ENDBSP}[1]{\BX{\ref{#1}}
                        \end{rm}\end{bespiel}
                       }

\newcommand{\prefix}{\sqsubseteq}
\newcommand{\backwd}[1]{\stackrel{\leftharpoonup}{#1}}
\newcommand{\down}{\;\downarrow\!}
\newcommand{\uniqueP}{\Upsilon}
\newcommand{\vzero}{0}
\newcommand{\from}{\leftarrow}
\newcommand{\drop}[1]{}

\renewcommand{\labelitemi}{$\bullet$}
\renewcommand{\labelitemii}{$\circ$}
\renewcommand{\labelitemiii}{$\cdot$}
\renewcommand{\labelitemiv}{$\ast$}

\newif\ifcomment
\commenttrue 
\definecolor{gris}{gray}{0.3}
\newcommand{\mycomment}[3]{\ifcomment
 {\small \null\-
  {\color{#1}{\textbf{#2:}} \color{#1} {#3}}}%
 \fi
}\newcommand{\EB}[1]{\mycomment{red!80!black}{EB}{#1}}
\newcommand{\RD}[1]{\mycomment{green!60!black}{RD}{#1}}
\newcommand{\US}[1]{\mycomment{blue!60!black}{US}{#1}}
\newcommand{\HW}[1]{\mycomment{green!60!black}{HW}{#1}}

\newcounter{exampleTScounter}
\newcommand{\TSref}[1]{\TS_{\ref{#1}}}
\newcommand{\PNref}[1]{\mathit{PNS}_{\ref{#1}}}
\newcommand{\PNS}{\mathit{PNS}}
\newcommand{\Nref}[1]{N_{\ref{#1}}}

\newcommand{\np}{\mathbb{N}_{\neq 0}}
\newcommand{\qp}{\mathbb{Q}_{>0}}

\renewcommand{\mod}{\text{ {\rm mod} }}

\newcommand{\CF}{\ensuremath{\mathrm{CF}}\xspace}
\newcommand{\EC}{\ensuremath{\mathrm{EC}}\xspace}
\newcommand{\EFC}{\ensuremath{\mathrm{EFC}}\xspace}
\newcommand{\WAC}{\ensuremath{\mathrm{WAC}}\xspace}
\newcommand{\ACC}{\ensuremath{\mathrm{AC}}\xspace}
\newcommand{\WRAC}{\ensuremath{\mathrm{WRAC}}\xspace}
\newcommand{\BRAC}{\ensuremath{\mathrm{BRAC}}\xspace}
\newcommand{\RAC}{\ensuremath{\mathrm{RAC}}\xspace}
\newcommand{\WPI}{\ensuremath{\mathrm{WCP}}\xspace}
\newcommand{\MG}{\ensuremath{\mathrm{MG}}\xspace}

\itemsep0pt

\title{Synthesis of Reduced Asymmetric Choice Petri Nets}

\author{Harro Wimmel\inst{1}\thanks{Supported by DFG (German Research Foundation)
through grant Be 1267/16-1 {\tt ASYST} (Algorithms for Synthesis and Pre-Synthesis Based on Petri Net Structure Theory).}
}

\institute{
Department of Computing Science,\\
Carl von Ossietzky Universit\"{a}t Oldenburg,
D-26111 Oldenburg, Germany \\
\email{harro.wimmel@informatik.uni-oldenburg.de}
}

\maketitle

\begin{abstract}
A Petri net is choice-free if any place has at most one transition in its postset (consuming its tokens) and
it is (extended) free-choice (\EFC) if the postsets of any two places are either equal or disjoint.
Asymmetric choice (\ACC) extends \EFC such that two places may also have postsets where one is contained in the other.
In reduced AC nets this containment is limited: If the postsets are neither disjoint nor equal, 
one is a singleton and the other has exactly two transitions. 
The aim of Petri net synthesis is to find an unlabelled Petri net in some target class with a reachability graph
isomorphic to a given finite labelled transition system (\lts).
Choice-free nets have strong properties, allowing to often easily detect when synthesis will fail or at least
to quicken the synthesis. With \EFC as the target class, only few properties can be checked ahead
and there seem to be no short cuts lowering the complexity of the synthesis (compared to arbitrary Petri nets).
For AC nets no synthesis procedure is known at all.
We show here how synthesis to a superclass of reduced AC nets (not containing the full AC net class) can be done.
\end{abstract}


\section{Introduction}
\label{intro.sct}

When dealing with the behaviour of Petri nets~\cite{reisig-2013,murata-89} there are two opposite approaches.
We can {\it analyse} a Petri net, building a variety of descriptions of its behaviour from sets of
firing sequences~\cite{hack-lang} to event structures~\cite{npw81}. One of the most common forms for describing the
sequential behaviour is the reachability graph, containing the reachable markings 
as nodes together with edges denoting transitions that fire to reach 
one marking from another.
In the reverse direction, i.e.\ {\it synthesis}~\cite{bbd15}, we can try to find a Petri 
net\footnote{we generally use Petri nets with arc weights} that 
behaves like a given specification, e.g. a labelled transition system (\lts).
Since using labelled Petri nets would always allow a trivial solution (isomorphic to the \lts),
we restrict ourselves to unlabelled Petri nets. As the reachability problem for
Petri nets is \EX-hard~\cite{lipton}, unlabelled nets can have a very complex behaviour, but on the
other hand there are even simple words, i.e.\ linearly ordered \lts, that are
not behaviours of such Petri nets~\cite{besw16}. This can make synthesis quite difficult;
even for the rather small \lts in the left of Fig.~\ref{f.gen1}, the resulting synthesised net on the right
is not immediately obvious.

\begin{figure}[t]
\centering
\begin{tikzpicture}[LTS]
\node[state,label=above:$s_0$] (s0) {};
\path[edge] (s0) ++(-0.5,0) edge (s0);
\node[state,below=of s0,label=left:$s_1$] (s1) {};
\node[state,right=of s1,label=below:$s_2$] (s2) {};
\node[state,right=of s0,label=above:$s_3$] (s3) {};
\node[state,below=of s1,label=left:$s_4$] (s4) {};
\node[state,right=of s2,label=below:$s_5$] (s5) {};
\node[state,right=of s4,label=above:$s_{10}$] (s10) {};
\node[state,below=of s10,label=left:$s_6$] (s6) {};
\node[state,right=of s5,label=above:$s_{12}$] (s12) {};
\node[state,below=of s12,label=right:$s_7$] (s7) {};
\node[state,above=of s5,label=above:$s_8$] (s8) {};
\node[state,right=of s6,label=right:$s_9$] (s9) {};
\node[state,right=of s10,label=below:$s_{11}$] (s11) {};
\node[state,right=of s12,label=below:$s_{13}$] (s13) {};
\node[state,above=of s13,label=above:$s_{14}$] (s14) {};
\draw[edge] (s0) edge node[auto,swap]{$a$} (s1);
\draw[edge] (s0) edge[bend left=10] node[auto]{$b$} (s3);
\draw[edge] (s1) edge node[auto,swap]{$d$} (s2);
\draw[edge] (s1) edge[bend left=10] node[auto]{$c$} (s4);
\draw[edge,densely dashed] (s2) edge node[auto,xshift=1mm,yshift=1mm]{$f$} (s0);
\draw[edge] (s2) edge node[auto]{$e$} (s5);
\draw[edge,densely dashed] (s3) edge[bend left=10] node[auto,xshift=1mm,yshift=0.5mm]{$c$} (s0);
\draw[edge,densely dashed] (s4) edge[bend left=10] node[auto]{$b$} (s1);
\draw[edge] (s4) edge node[auto,swap]{$a$} (s6);
\draw[edge] (s5) edge node[auto,xshift=-0.8mm,yshift=-0.8mm]{$a$} (s7);
\draw[edge] (s5) edge[bend left=10] node[auto]{$b$} (s8);
\draw[edge] (s6) edge[bend left=10] node[auto]{$c$} (s9);
\draw[edge] (s6) edge node[auto,swap]{$d$} (s10);
\draw[edge,densely dashed] (s7) edge[bend left=10] node[auto]{$c$} (s11);
\draw[edge] (s7) edge node[auto,swap]{$d$} (s12);
\draw[edge,densely dashed] (s8) edge[bend left=10] node[auto]{$c$} (s5);
\draw[edge,densely dashed] (s9) edge[bend left=10] node[auto]{$b$} (s6);
\draw[edge,densely dashed] (s10) edge node[auto,swap]{$f$} (s4);
\draw[edge] (s10) edge node[auto]{$e$} (s11);
\draw[edge,densely dashed] (s11) edge[bend left=10] node[auto,xshift=-1mm,yshift=-0.5mm]{$b$} (s7);
\draw[edge,densely dashed] (s12) edge node[auto,swap]{$f$} (s5);
\draw[edge] (s12) edge node[auto]{$e$} (s13);
\draw[edge] (s13) edge[bend left=10] node[auto]{$b$} (s14);
\draw[edge,densely dashed] (s14) edge[bend left=10] node[auto]{$c$} (s13);
\end{tikzpicture}\hspace*{1.5cm}
\begin{tikzpicture}
\draw(0.3,3) node[place,tokens=2,label=left:$p_1$] (p1) {};
\draw(1.7,3) node[place,tokens=1,label=left:$p_2$] (p2) {};
\draw(0,1.5) node[transition] (a) {$a$};
\draw(1,1.5) node[transition] (b) {$b$};
\draw(2,1.5) node[transition] (c) {$c$};
\draw(0.3,0) node[place,label=left:$p_3$] (p3) {};
\draw(1.7,0) node[place,label=left:$p_4$] (p4) {};
\draw(3.5,0) node[transition] (d) {$d$};
\draw(3.5,1.5) node[place,label=right:$p_5$] (p5) {};
\draw(3,3) node[transition] (e) {$e$};
\draw(4,3) node[transition] (f) {$f$};
\draw(p1) edge[-latex,thick] (a);
\draw(p2) edge[-latex,thick] (a);
\draw(a) edge[-latex,thick] (p3);
\draw(a) edge[-latex,thick] (p4);
\draw(p2) edge[-latex,thick] (b);
\draw(b) edge[-latex,thick] (p4);
\draw(p4) edge[-latex,thick] (c);
\draw(c) edge[-latex,thick] (p2);
\draw(p3) edge[-latex,thick,bend right=30] (d);
\draw(p4) edge[-latex,thick] (d);
\draw(d) edge[-latex,thick] (p5);
\draw(p5) edge[-latex,thick] (e);
\draw(e) edge[-latex,thick] (p2);
\draw(p5) edge[-latex,thick] (f);
\draw(f) edge[-latex,thick,bend right=35] (p1);
\draw(f) edge[-latex,thick,bend right=25] (p2);
\end{tikzpicture}
\caption{An \lts (left) that can be synthesised to a reduced asymmetric choice Petri net (right).
Solid edges in this \lts form a spanning tree $E$, other edges (`chords') are dashed.
States have Parikh vectors according to their spanning tree walks, e.g.\ $\parikh_E(s_9)=\parikh(acac)=2a+2c$ since $s_0\step{acac}_Es_9$.
The Parikh vector of the chord $s_7\step{c}s_{11}$ (with the spanning tree walks $s_0\step{adea}_Es_7$ and
$s_0\step{acade}_Es_{11}$) is $\parikh_E(s_7\step{c}s_{11})=\parikh_E(s_{7})+1c-\parikh_E(s_{11})=(2a+1d+1e)+1c-(2a+1c+1d+1e)=\zero$.
All other chords have Parikh vectors $1b+1c$ or $1a+1d+1f$}
\label{f.gen1}
\end{figure}
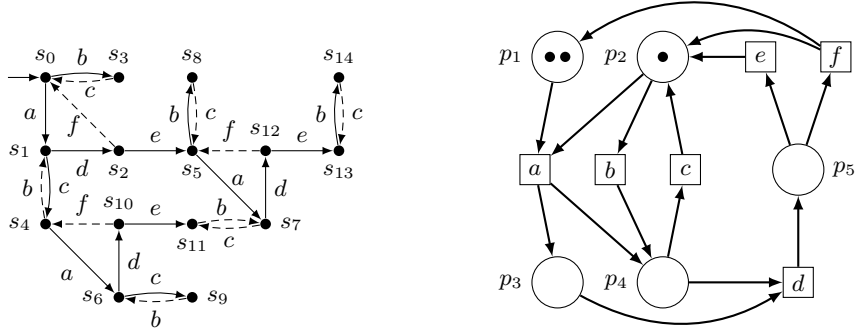

Region theory~\cite{er89,bd99} connects states of an \lts with markings of a Petri net
and determines two kinds of problems that need to be solved
for a successful synthesis. {\it State separation problems} demand that distinguished
states in the \lts must correspond to different markings in a Petri net. {\it Event-state
separation problems} enforce the non-firability of a Petri net transition in a marking
if the \lts has no outgoing edge with the matching label in the corresponding state.
Both kinds of problems can be formulated as linear inequality systems~\cite{bds16,sw18} and tackled
via e.g.\ SMT-solvers. Seeking solutions in the integer domain (as generally required for Petri nets)
is \NP-complete, but often rational solutions are sufficient, 
and for this case Karmarkar~\cite{karmarkar} provided an algorithm with a polynomial worst case complexity.
Generally, if all constant terms in the linear inequality systems are zero, we can use rational solutions
and multiply them by appropiate factors to lift them to integers.
E.g.,\ the inequality system $x+1\le y\le x+y\le 2\le 4x$ with the contant term $2$ has the rational solution
$x=0.5$, $y=1.5$, but no integer solutions.

Often, we would like to find a Petri net with some additional properties, i.e.\ we target for
some subclass of Petri nets (see also Fig.~\ref{f.classes}). The most common case is the class of bounded Petri nets, which are
exactly the nets with finite \lts as behaviours, allowing us to work directly with the \lts as
input. Very limited classes like marked graphs (\MG)~\cite{bd17b} and choice-free nets (\CF)~\cite{bds17} 
have been investigated to determine whether structural analysis of an \lts allows to reduce
the size of the linear inequality systems to be solved for synthesis, or even forego them altogether.
Some overview of properties of Petri nets has been done~\cite{s16}, with the result that some subclasses can
easily be targetted (with canonical approaches that are also combinable), while other simple properties
cannot be tackled at all. For the subclass \EC of bounded equal-conflict Petri nets (where the postsets of places 
are either identical up to arc weights or completely disjoint), a structural analysis 
of the \lts is necessary first~\cite{sp17}. This analysis either
determines that the synthesis must fail or it provides the modifications required for the
linear inequality systems (in which case the synthesis may still fail). These modifications enforce
the result to be an equal-conflict net, if successful. The well-known (extended) free-choice nets (\EFC)~\cite{bes87} are the subclass
of equal-conflict nets where arc weights are limited to one (i.e.\ plain equal-conflict nets).
Equal-conflict nets can be synthesised in polynomial worst case complexity. For 
free-choice synthesis no such result is known due to the limit on arc weights,
but at least an algorithm in \NP{} exists.

\begin{figure}[t]
\centering
\begin{tikzpicture}
\draw(0,0) circle [radius=4mm];
\node(0,0) {\MG};
\draw[thick](0,-1.0) ellipse [x radius=8mm,y radius=1.7cm];
\draw(0.0,-2.5) node {\CF};
\draw(0,0) circle [radius=1.2cm];
\draw(0.0,1.0) node {\EFC};
\draw[thick](0,-0.75) circle [radius=2cm];
\draw(1.0,-2.2) node {\EC};
\draw(0,0) circle [radius=1.6cm];
\draw(0.0,1.4) node {\RAC};
\draw(0,0) circle [radius=2cm];
\draw(0.0,1.8) node {\BRAC};
\draw(-0.5,0) ellipse [x radius=3.2cm,y radius=2.2cm];
\draw(-3.4,0.0) node {\ACC};
\draw(-0.6,-0.25) ellipse [x radius=3.7cm,y radius=2.65cm];
\draw(-3.3,-1.6) node {\WAC};
\draw[thick](0.6,-0.25) ellipse [x radius=3.7cm,y radius=2.65cm];
\draw(3.8,-0.25) node {\WPI};
\draw(0.0,-0.1) ellipse [x radius=4.7cm,y radius=3.0cm];
\draw(0,2.7) node {bounded nets};
\end{tikzpicture}
\caption{Relations between classes of Petri nets relevant for our work: marked graph (\MG), choice-free (\CF), 
extended free-choice (\EFC), reduced asymmetric choice (\RAC), block-reduced asymmetric choice (\BRAC), 
asymmetric choice (\ACC), equal-conflict (\EC), asymmetric choice with arc weights (\WAC), and weighted comparable preset (\WPI).
The last three classes and \CF allow arc weights greater than one 
\label{f.classes}}
\end{figure}
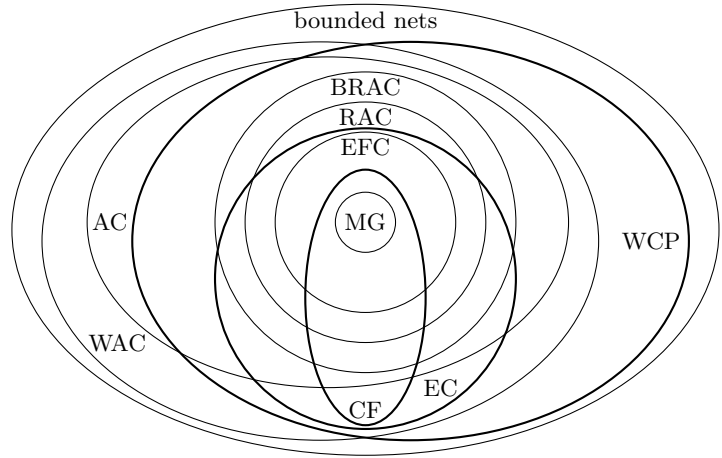

For free-choice nets, structural and behavioural properties are strongly connected.
Commoner's Theorem and the Rank Theorem~\cite{de95} are two well-known examples linking
liveness and well-formedness of a net to siphons/marked traps and the rank of the net's
incidence matrix, respectively.
In modern research areas like business processes and web services, these properties play
an important role, but the modelled systems are seldom free-choice.
Asymmetric choice nets (\ACC)~\cite{commoner,bes87} are an extension of free-choice nets
with a broader applicability where the important theorems still hold at least 
partially~\cite{de95,jcl04} and their properties have also been investigated with
respect to complexity issues~\cite{lj15}. \ACC nets allow confusion to happen, 
an asymmetric combination of choice and concurrency. The postsets of any two places 
may be disjoint or identical (like for free-choice) or one postset may
be contained in the other. Extensions have been defined~\cite{akd98}, but in this paper we will
be more interested in restrictions, i.e.\ classes between free-choice and
asymmetric choice, in the hope of finding some class where synthesis
is possible.

Compared to \ACC,
reduced asymmetric choice (\RAC)~\cite{commoner} limits the number of transitions in the postsets of two places
that are properly contained in one another: one postset must be a singleton, the other contains exactly two transitions.
Fig.~\ref{f.gen1} shows an example where the postset of $p_1$ is contained in that of $p_2$
(written as $W(p_1,\cdot)=1a\le 1a+1b=W(p_2,\cdot)$ or $p_1{}^\bullet=\{a\}\subseteq\{a,b\}=p_2{}^\bullet$). A similar relation
holds for $p_3$ and $p_4$ while $p_5$ is independent and presents a free choice between $e$ and $f$.
We will also look at classes with arc weights. Figure~\ref{f.rac} shows some example nets from
these classes. In this paper, we will synthesise \lts targetting at some superclasses of reduced asymmetric
choice nets, but not the full class of asymmetric choice nets. As shown in~\cite{bes87},
there are strong relations between the classes of asymmetric choice and reduced asymmetric choice
net via a marking simulation, but in this paper we rather look at \lts and thus labelled firing sequences.

\begin{figure}[t]
\centering
\begin{tikzpicture}
\draw(0,0) node[transition] (t1) {$t_1$};
\draw(1,0) node[transition] (t2) {$t_2$};
\draw(2,0) node[transition] (t3) {$t_3$};
\draw(0,2) node[place,label=above:$p_1$] (p1) {};
\draw(1,2) node[place,label=above:$p_2$] (p2) {};
\draw(2,2) node[place,label=above:$p_3$] (p3) {};
\draw(p1) edge[-latex,thick] node[auto,swap]{$2$} (t1);
\draw(p2) edge[-latex,thick] node[auto,swap]{$2$} (t2);
\draw(p2) edge[-latex,thick] node[auto,swap,near end,xshift=-1mm,yshift=1mm]{$2$} (t3);
\draw(p3) edge[-latex,thick] node[auto,swap,near start,xshift=-1mm]{$3$} (t2);
\draw(p3) edge[-latex,thick] node[auto]{$3$} (t3);
\end{tikzpicture}\hspace*{1cm}
\begin{tikzpicture}
\draw(0,0) node[transition] (t1) {$t_1$};
\draw(1,0) node[transition] (t2) {$t_2$};
\draw(2,0) node[transition] (t3) {$t_3$};
\draw(0,2) node[place,label=above:$p_1$] (p1) {};
\draw(1,2) node[place,label=above:$p_2$] (p2) {};
\draw(2,2) node[place,label=above:$p_3$] (p3) {};
\draw(p1) edge[-latex,thick] node[auto,swap]{$2$} (t1);
\draw(p2) edge[-latex,thick] node[auto,swap,xshift=1mm]{$2$} (t1);
\draw(p2) edge[-latex,thick] (t2);
\draw(p3) edge[-latex,thick] node[auto,near end,xshift=-2mm]{$2$} (t1);
\draw(p3) edge[-latex,thick] node[auto,xshift=-1mm]{$2$} (t2);
\draw(p3) edge[-latex,thick] node[auto]{$3$} (t3);
\end{tikzpicture}\hspace*{1cm}
\begin{tikzpicture}
\draw(0,0) node[transition] (t1) {$t_1$};
\draw(1,0) node[transition] (t2) {$t_2$};
\draw(2,0) node[transition] (t3) {$t_3$};
\draw(0,2) node[place,label=above:$p_1$] (p1) {};
\draw(1,2) node[place,label=above:$p_2$] (p2) {};
\draw(2,2) node[place,label=above:$p_3$] (p3) {};
\draw(p1) edge[-latex,thick] node[auto,swap]{$2$} (t1);
\draw(p2) edge[-latex,thick] node[auto,swap]{$4$} (t2);
\draw(p2) edge[-latex,thick] node[auto,xshift=-1mm]{$3$} (t3);
\draw(p3) edge[-latex,thick] node[auto]{$2$} (t3);
\end{tikzpicture}
\caption{Left: an equal-conflict net, $t_2$ and $t_3$ are in conflict and 
both are concurrent to $t_1$. Middle: a weighted asymmetric choice net where the three places have increasing
postsets with $W(p_1,\cdot)\lneqq W(p_2,\cdot)\lneqq W(p_3,\cdot)$. Right: a weighted reduced asymmetric choice net, only structures forming an `N' are allowed
for asymmetric choices, $W(p_3,\cdot)=2t_3\lneqq 4t_2+3t_3=W(p_2,\cdot)$
\label{f.rac}}
\end{figure}
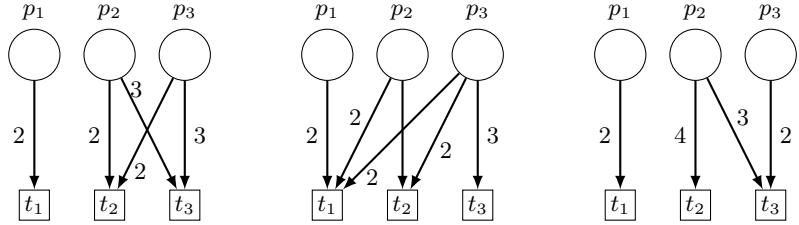

In the next section, we will introduce the basic concepts around labelled transition systems and
Petri nets as well as a short description of synthesis and how separation problems are defined.
Section~\ref{sect.3} draws conclusions from an \lts for a synthesised net when we target one of two specific
classes. The first class, {\em block-reduced asymmetric choice} (\BRAC), lies between asymmetric choice
and reduced asymmetric choice. The other class (\WPI) is a superclass of \BRAC which lies askew to
asymmetric choice nets. In \WPI (weighted comparable presets) the presets of transitions 
are either disjoint or comparable and we allow arbitrary arc weights.
One important conclusion is that self-loops (edges with the same source and target node) in the \lts make synthesis intrinsically
difficult. In Section~\ref{sect.4} we investigate the problems that stem from self-loops for the class \WPI,
and in Section~\ref{sect.5} we solve these problems for the class \BRAC.
Finally, we give a summary and an outlook in Section~\ref{sect.O}.

\section{Basic concepts}
\label{basic-not.sct}

\DEF{d.lts}{LTS}
A {\it labelled transition system} (\lts) with initial state is a tuple $TS=(S,\Sigma,\to,s_0)$ 
with nodes $S$ (a countable set of states),
edge labels $\Sigma$ (a finite set of letters),
edges $\mathord{\to}\subseteq(S\times \Sigma\times S)$,
and an initial state $s_0\in S$. An edge $(s,t,s')\in\mathord{\to}$ may be written as
$s\step{t}s'$.  
A {\it walk} $\sigma\in \Sigma^*$ from $s$ to $s'$, written as
$s\step{\sigma}s'$, 
is given inductively by $s=s'$ for the empty word $\sigma=\varepsilon$ and by
$\exists s''\in S$: $s\step{w}s''\step{t}s'$ for $\sigma=wt$ 
with $w\in \Sigma^*$ and $t\in \Sigma$. A walk $s\step{\sigma}s'$ is a {\it cycle} 
if and only if $s=s'$, we call it a {\it self-loop} if $\sigma\in\Sigma$.
The set $\step{s}$ for $s\in S$ is the set of all states reachable
from $s$, $\step{s} = \{s'\,|\,\exists \sigma\in \Sigma^*:\;s\step{\sigma}s'\}$.
The {\it Parikh vector} $\parikh(\sigma)\colon\Sigma\to\Z$ of a word $\sigma\in \Sigma^*$ maps each letter $t\in \Sigma$ 
to its number of occurrences in $\sigma$, it will often be written as an element of the
group spanned by $\Sigma$. The neutral element is written as \zero, comparisons are done componentwise
with $\mathord{\lneqq}$ meaning ``less or equal in all components, but not entirely equal''. 
We map to $\Z$ here instead of $\N$ to be able to extend the notion of a Parikh vector later and
to handle differences of Parikh vectors more easily.

A {\it spanning tree} $E$ of $TS$ is a set of edges $E\subseteq\mathord{\to}$ 
such that for every $s\in S$ there is a unique walk from $s_0$ to $s$ using edges in $E$ only.
This implies that $E$ is cycle-free. 
A {\it walk in $E$} is a walk that uses edges in $E$ only (and not any of $\mathord{\to}\backslash E$).
Edges in $\mathord{\to}\backslash E$ are called {\it chords}.
The {\it Parikh vector of a state} $s$ in a spanning tree $E$ is $\parikh_E(s) = \parikh(\sigma)$
where $s_0\step{\sigma}s$ is the unique walk in $E$. The {\it Parikh vector of an edge}
$s\step{t}s'$ in $TS$ is $\parikh_E(s\step{t}s') = \parikh_E(s)+1t-\parikh_E(s')$, see
Fig.~\ref{f.gen1} for an example. 
Note that Parikh vectors of edges in $E$ always evaluate to zero; for chords the Parikh vector may even 
contain negative values. For a chord $s\step{t}s'$, $s$ and $s'$ have a latest common predecessor $r$ 
in $E$, $t',t''\in T$ with $t'\not=t''$, $\sigma,\sigma'\in T^*$ with two walks
$r\step{t'\sigma}s\step{t}s'$ and $r\step{t''\sigma'}s'$ in $E$. These two walks form a {\it cycle in the
LTS' underlying undirected graph}. If we follow the cycle in the direction of the chord
and sum up the edges, we obtain the chord's Parikh vector $\parikh_E(s)+1t-\parikh_E(s')$. 
The {\it Parikh vector of a walk} $s_1\step{t_1}s_2\ldots s_n\step{t_n}s_{n+1}$ is defined as 
$\parikh_E(s_1\step{t_1\ldots t_n}s_{n+1}) = \sum_{i=1}^n\parikh_E(s_i\step{t_i}s_{i+1})$.
Obviously, $\parikh_E(s_1\step{t_1\ldots t_n}s_{n+1}) = \parikh_E(s_1)+\parikh(t_1\ldots t_n)-\parikh_E(s_{n+1})$.
If the walk is a cycle (with $s_1=s_{n+1}$), we thus find 
$\parikh(t_1\ldots t_n)=\sum_{i=1}^n\parikh_E(s_i\step{t_i}s_{i+1})$ where all non-zero
Parikh vectors in the sum stem from chords. 
The set $\{\parikh_E(s\step{t}s')\,|\,(s,t,s') \in \mathord{\to}\backslash E\}$ is then a generator
for all Parikh vectors of cycles (the latter being linear combinations of its elements).
By simple linear algebra, 
we can compute a basis from this generator. This {\it cycle base} $\Gamma$ contains
at most $|\Gamma|\le|\Sigma|$ different Parikh vectors.
There will be no need to distinguish between cycles in the \lts and in its underlying undirected
graph at all.

An \lts $TS=(S,\Sigma,\to,s_0)$ is {\it finite} if $S$ is finite and it is {\it deterministic} 
if $s\step{t}s'$ and $s\step{t}s''$ implies $s'=s''$ for all $s\in S$ and $t\in \Sigma$. We call
$TS$ {\it reachable} if for every state $s\in S$ exists some $\sigma\in \Sigma^*$ with $s_0\step{\sigma}s$.
Reachability implies the existence of a spanning tree. 
Two labelled transition systems $TS_1=(S_1,\Sigma_1,\to_1,s_{01})$ and $TS_2=(S_2,\Sigma_2$, $\to_2,s_{02})$ are
{\it isomorphic} if $\Sigma_1=\Sigma_2$ and there is a bijection $\zeta\colon S_1\to S_2$ with $\zeta(s_{01})=s_{02}$ and
$(s,t,s')\in\;\to_1\,\Leftrightarrow(\zeta(s),t,\zeta(s'))\in\;\to_2$, for all $s,s'\in S_1$.
\ENDDEF{d.lts}

An example for a spanning tree $E$ and some Parikh vectors of states and chords is shown in Fig.\ref{f.gen1}.

\DEF{d.pn}{Petri nets} 
An {\em (initially marked) Petri net} is denoted as $N=(P,T,W,M_0)$ where $P$ is a finite set of places,
$T$ is a finite set of transitions,
$W$ is the weight function $W\colon((P\times T)\cup(T\times P))\to\N$ 
specifying the arc weights,
and $M_0$ is the initial marking
(where a marking is a mapping $M\colon P\to\N$, indicating the number of tokens in each place).
The {\em preset} of a place or transition $x\in P\cup T$ is defined as ${}^\bullet x = \{y\mid W(y,x)>0\}$
and its {\em postset} is $x^\bullet = \{y\mid W(x,y)>0\}$. We canonically extend this notion to
${}^\bullet X$ and $X^\bullet$ for sets $X\subseteq P\cup T$.
A transition $t\in T$ is {\it enabled} at a marking $M$,
denoted by $M\step{t}$, if $\forall p\in P\colon M(p)\geq W(p,t)$.
The {\it firing} of $t$ leads from $M$ to $M'$, denoted by $M\step{t}M'$,
if $M\step{t}$ and $M'(p)=M(p)-W(p,t)+W(t,p)$.
This can be extended, by induction as usual, to $M\step{\sigma}M'$ for words $\sigma\in T^*$,
and $\step{M} = \{M'\,|\,\exists\sigma\in T^*\colon M\step{\sigma}M'\}$
denotes the set of markings reachable from $M$.
The {\it reachability graph} $RG(N)$ of a Petri net $N$
is the labelled transition system with the set of nodes $\step{M_0}$,
initial state $M_0$, label set $T$,
and set of edges $\{(M,t,M')\mid M,M'\in\step{M_0}\land M\step{t}M'\}$.
If a labelled transition system $TS$ is isomorphic to the reachability graph $RG(N)$ of 
a Petri net $N$ we say that $N$ {\em PN-solves} (or simply {\em solves}) $TS$,
and that $TS$ is {\em synthesisable} to $N$. 

A Petri net $N$ is {\it bounded} if there is some $k\in\N$ such that $\forall p\in P$ $\forall M\in\step{M_0}$: $M(p)\le k$,
which is equivalent to its reachability graph being finite.
We define some subclasses of bounded Petri nets by certain properties.
A Petri net $(P,T,W,M_0)$ is
\begin{itemize}
\item {\em plain} if $W\colon (P\times T)\cup(T\times P)\to \{0,1\}$.
\item {\em choice-free} (\CF) if $\forall p\in P, t,t'\in T\colon W(p,t)>0 \wedge t\neq t' \then W(p,t')=0$.
\item {\em equal-conflict} (\EC) if $\forall t,t'\in T\colon {}^\bullet t\cap {}^\bullet t'\neq\emptyset \then W(\cdot,t)=W(\cdot,t')$.\footnote{$W(\cdot,t)\colon P\to\N$ is the submapping of $W$ with fixed second parameter $t$} 
\item {\em (extended) free-choice} (\EFC) if it is \EC and plain.
\item {\em weighted comparable preset} (\WPI) if $\forall t,t'\in T\colon {}^\bullet t\cap {}^\bullet t'\neq\emptyset \then W(\cdot,t)\geq W(\cdot,t') \vee W(\cdot,t)\leq W(\cdot,t')$.\footnote{Comparisons of mappings are done componentwise, cf.\ Parikh vectors}
\item {\em weighted asymmetric choice} (\WAC) if $\forall p,p'\in P\colon p^\bullet\cap p'{}^\bullet\neq\emptyset \then W(p,\cdot)\geq W(p',\cdot) \vee W(p,\cdot)\leq W(p',\cdot)$.
\item {\em asymmetric choice} (\ACC) if it is \WAC and plain. 
\item {\em reduced asymmetric choice} (\RAC) if it is plain and $\forall p,p'\in P\colon p^\bullet\cap p'{}^\bullet\neq\emptyset \then (|p^\bullet|=1\wedge|p'{}^\bullet|\le 2\wedge {}^\bullet(p'{}^\bullet)=\{p,p'\})\vee(|p'{}^\bullet|=1\wedge|p{}^\bullet|\le 2\wedge {}^\bullet(p{}^\bullet)=\{p,p'\})$.
\item {\em block-reduced asymmetric choice} (\BRAC) if it is plain and $\forall p,p'\in P\colon p^\bullet\cap p'{}^\bullet\neq\emptyset \then (p^\bullet=p'{}^\bullet \vee \exists T_1,T_2\subseteq T\colon (p^\bullet=T_1\wedge p'{}^\bullet=T_1\cup T_2\wedge {}^\bullet T_1=\{p,p'\}\wedge {}^\bullet T_2=\{p'\}) \vee (p'{}^\bullet=T_1\wedge p^\bullet=T_1\cup T_2\wedge {}^\bullet T_1=\{p,p'\}\wedge {}^\bullet T_2=\{p\}))$.
\end{itemize}
All property names are also used as class names, e.g.\ AC is the class of all bounded asymmetric choice Petri nets.
\ENDDEF{d.pn}

Note that the free-choice property \EFC can be alternatively defined via
$\forall p,p'\in P\colon p^\bullet \cap p'{}^\bullet \neq\emptyset \then W(p,\cdot)=W(p',\cdot)$
(plus plainness), but for \EC a place-based definition is not possible.
In \EC, all transitions consuming tokens from a place must take the same amount,
while the place-based alternative definition would demand that a transition must
take the same number of tokens from each place in its preset.
As a consequence, the classes \WPI and \WAC are not identical and have \EFC in 
their intersection, but not \EC. The right net in Fig.~\ref{f.rac} is in \WAC
but not in \WPI as $W(p_2,t_3)\lneqq W(p_2,t_2)$. If we swap arc weights, setting
$W(p_2,t_2)=2$ and $W(p_3,t_3)=4$, the net is in \WPI but not in \WAC.

The definition of \RAC stems from~\cite{commoner} and can also be found in~\cite{bes87}. A transition that does not share its preset with another
transition may have an arbitrary preset (with $|p^\bullet|=1$ and $|p'{}^\bullet|=1$).
Other transitions (in the postset of $p$/$p'$ with $|p^\bullet|=2$ or $|p'{}^\bullet|=2$) may share
their preset with at most one other transition. One of $p$/$p'$ will be in the preset of only one
of these transitions, the other in both, forming the known `N'-structure in the Petri net. For transitions
with identical presets, this preset may only contain a single place.

\BRAC extends this (see Fig.~\ref{f.brac}) by allowing {\em blocks of transitions} with the same preset taking the place of
single transitions in \RAC. An {\em asymmetric choice block} consists of two separate blocks of transitions, here 
$T_1=\{t_6,t_7,t_8\}$ and $T_2=\{t_4,t_5\}$, 
where all transitions in $T_2$ have the same (single) place ($p_4$) in their presets. 
The presets in the other block ($T_1$) contain the same additional place ($p_5$).
If we have two places (here $p_4$, $p_5$) with non-identical postsets but sharing a transition,
the condition ${}^\bullet T_1=\{p,p'\}$ demands that only those two places can be involved. Once
a third place comes into play (as with $\{p_1,p_2,p_3\})$ all places must have identical postsets,
forming a {\em free-choice block}.

\begin{figure}[t]
\centering
\begin{tikzpicture}
\draw(0,0) node[transition] (t1) {$t_1$};
\draw(1,0) node[transition] (t2) {$t_2$};
\draw(2,0) node[transition] (t3) {$t_3$};
\draw(4,0) node[transition] (t4) {$t_4$};
\draw(5,0) node[transition] (t5) {$t_5$};
\draw(6,0) node[transition] (t6) {$t_6$};
\draw(7,0) node[transition] (t7) {$t_7$};
\draw(8,0) node[transition] (t8) {$t_8$};
\draw(0,2) node[place,label=above:$p_1$] (p1) {};
\draw(1,2) node[place,label=above:$p_2$] (p2) {};
\draw(2,2) node[place,label=above:$p_3$] (p3) {};
\draw(5,2) node[place,label=above:$p_4$] (p4) {};
\draw(7,2) node[place,label=above:$p_5$] (p5) {};
\draw(p1) edge[-latex,thick] (t1);
\draw(p1) edge[-latex,thick] (t2);
\draw(p1) edge[-latex,thick] (t3);
\draw(p2) edge[-latex,thick] (t1);
\draw(p2) edge[-latex,thick] (t2);
\draw(p2) edge[-latex,thick] (t3);
\draw(p3) edge[-latex,thick] (t1);
\draw(p3) edge[-latex,thick] (t2);
\draw(p3) edge[-latex,thick] (t3);
\draw(p4) edge[-latex,thick] (t4);
\draw(p4) edge[-latex,thick] (t5);
\draw(p4) edge[-latex,thick] (t6);
\draw(p4) edge[-latex,thick] (t7);
\draw(p4) edge[-latex,thick] (t8);
\draw(p5) edge[-latex,thick] (t6);
\draw(p5) edge[-latex,thick] (t7);
\draw(p5) edge[-latex,thick] (t8);
\end{tikzpicture}
\caption{The two kinds of preset structures allowed in block-reduced asymmetric choice nets:
either extended free-choice with arbitrarily large presets (left) or two blocks of transitions
forming asymmetric choices where all transitions in one block have the same preset of either one
or two places (right)
\label{f.brac}}
\end{figure}
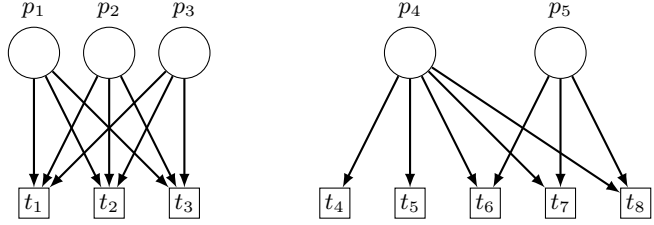

\KOR{c.pn}{Class Inclusions}
We get some simple inclusions, also shown in Fig.~\ref{f.classes}:\pagebreak

\begin{itemize}
\item $\EFC \subseteq \RAC \subseteq \BRAC \subseteq \ACC \subseteq \WAC$,
\item $\CF \subseteq \EC \subseteq \WAC$,
\item $\EFC \subseteq \EC$, and
\item $\BRAC \subseteq \WPI$.
\end{itemize}
\ENDKOR{c.pn}

Most of the inclusions are quite obvious, for $\RAC\subseteq\BRAC$ the condition
$|p'{}^\bullet|\le 2$ is split such that the case $|p'{}^\bullet|=1$ implies $p^\bullet=p'{}^\bullet$ 
while $|p'{}^\bullet|=2$ translates to $p'{}^\bullet=T_1\cup T_2$ in the \BRAC definition,
with $T_1$ and $T_2$ being singleton sets.
For $\BRAC\subseteq\WPI$, a place $p\in {}^\bullet t\cap {}^\bullet t'$ is either the only place in these
presets or the \BRAC-condition holds. In both cases, the presets are contained in one another.

We will mainly be interested in \WPI and \BRAC, but the results we show also hold for classes $C$
with $\BRAC\subseteq C\subseteq \WPI\cap\WAC$, which includes classes with arc weights. Essentially,
we can have arbitrary arc weights in the postsets of transitions and for presets of transitions
that fall under the \EC-condition. Whenever a transition takes part in an asymmetric choice,
the arc weights in its presets are mostly limited to one.

\DEF{d.synth}{Synthesis~\cite{bbd15}}
A {\it region} $r=(R,B,F)$ of an LTS $(S,\Sigma,\to,s_0)$ consists of three functions
$R$: $S\to\N$, $B$: $\Sigma\to\N$, and $F$: $\Sigma\to\N$ such that for all edges $s\step{t}s'$
in the LTS we have $R(s)\ge B(t)$ and $R(s')=R(s)-B(t)+F(t)$. This mimics the firing
rule of Petri nets and makes regions essentially equivalent to places, i.e.\ a place
$p$ can be defined from $r$ via $M_0(p)=R(s_0)$, $W(p,t)=B(t)$, and $W(t,p)=F(t)$ for
all $t\in \Sigma$. When a Petri net is constructed from a set of regions
of a reachable LTS,
this implies
a {\it uniquely defined marking} $\M(s)$ for each state $s$ with 
$\M(s)(p)=M_0(p)+\sum_{t\in \Sigma}\parikh(\sigma)(t)\cdot(W(t,p)-W(p,t))$
for an arbitrary walk $s_0\step{\sigma}s$.

The construction of a Petri net $N=(P,T,W,M_0)$ with one place in $P$
for each region of the LTS guarantees $s\step{t} \then \M(s)\step{t}$,
but has three issues: (1) $P$ might become infinite, (2) $\M$ may not be injective,
and (3) $\M(s)\step{t} \then s\step{t}$ need not hold. Failing (2) means
that there are states $s,s'\in S$ with $s\not=s'$ that are identified in $RG(N)$ (leading to non-isomorphism).
A {\it state separation problem} (SSP) is a pair $(s,s')\in S\times S$ with $s\not=s'$.
A region $r$ solves an SSP $(s,s')$ if $R(s)\not=R(s')$ (and thus $\M(s)\not=\M(s')$).
An example of an SSP in Fig.~\ref{f.genx} is $(s_3,s_7)$, for which no solving region exists
due to the fact that $R(s_3)=R(s_4)-B(b)+F(b)-B(a)+F(a)=R(s_4)-B(b)+F(b)-B(a)+F(a)=R(s_7)$.
Failing (3) results in an edge $\M(s)\step{t}$ in $RG(N)$ but not in the LTS,
$\neg s\step{t}$.
An {\it event/state separation problem} (ESSP) is a pair $(s,t)\in S\times \Sigma$
with $\neg s\step{t}$. A region $r$ solves an ESSP $(s,t)$ if $R(s)<B(t)$
(and thus $\neg\M(s)\step{t}$). 
An example is the ESSP $(s_2,b)$ in Fig.~\ref{f.genx}, which is unsolvable due to
the fact that $R(s_0)\ge B(b)$ (as $s_0\step{b}$), $R(s_2)=R(s_0)-B(a)+F(a)<B(b)$ (as $\neg s_2\step{b}$),
and $R(s_4)=R(s_2)-B(a)+F(a)\ge B(b)$ (as $s_4\step{b}$). Then, $R(s_0)\le R(s_2)\le R(s_4)$ or
$R(s_0)\ge R(s_2)\ge R(s_4)$, contradicting one of the comparisons to $B(b)$.
The set of all {\it separation problems}, $\{(s,s')\in S\times S\,|\,s\not=s'\}\cup\{(s,t)\in S\times \Sigma \mid \neg s\step{t}\}$,  
is finite for finite LTS, and finding
a solution for every separation problem solves all three issues, making $RG(N)$
isomorphic to the finite LTS, thus making the latter synthesisable.

\begin{figure}[t]
\centering
\begin{tikzpicture}[LTS]
\node[state,label=below:$s_0$] (s0) {};
\path[edge] (s0) ++(-0.5,0) edge (s0);
\node[state,above=of s0,label=above:$s_1$] (s1) {};
\foreach \x/\y in {0/2,2/4,4/6} {
 \node[state,right=of s\x,label=below:$s_{\y}$] (s\y) {}; }
\foreach \x/\y in {1/3,3/5,5/7} {
 \node[state,right=of s\x,label=above:$s_{\y}$] (s\y) {}; }
\draw[edge] (s0) edge node[auto,swap]{$b$} (s1);
\draw[edge] (s0) edge node[auto]{$a$} (s2);
\draw[edge] (s2) edge node[auto]{$a$} (s4);
\draw[edge] (s4) edge node[auto]{$a$} (s6);
\draw[edge] (s4) edge node[auto]{$b$} (s5);
\draw[edge] (s5) edge node[auto]{$a$} (s3);
\draw[edge] (s3) edge node[auto]{$b$} (s2);
\draw[edge] (s6) edge node[auto]{$b$} (s7);
\end{tikzpicture}
\caption{An LTS with unsolvable SSP $(s_3,s_7)$ and ESSP $(s_2,b)$}
\label{f.genx}
\end{figure}
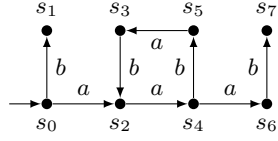

An SSP $(s,s')$ can be written as a linear inequality system~\cite{bds16}
with $\forall\gamma\in\Gamma$: $(F-B)\tp\cdot\gamma = 0$ (all cycles must have
effect zero in a region), $(F-B)\tp\cdot(\parikh_E(s)-\parikh_E(s')) \neq 0$
($\M(s)\neq\M(s')$, compared via $R(s_0)$), 
and $\forall s''\step{t}s'''$: $R(s_0)+(F-B)\tp\cdot\parikh_E(s'')\ge B(t)$
(definition of a region, but without checking cycles).
Here $F\ge\zero$, $B\ge\zero$ and $R(s_0)\ge 0$ are the variables, 
a solution forming a region solving the SSP. Note that if the
first two formulas can be solved, $R(s_0)$ can always be selected 
high enough such that the third one is also fulfilled. 

An ESSP $(s,t)$ can be written as a linear inequality system~\cite{sw18} with
$\forall\gamma\in\Gamma$: $(F-B)\tp\cdot\gamma = 0$,
$\forall s'\step{t'}s''$: $R(s_0)+(F-B)\tp\cdot\parikh_E(s')\ge B(t')$,
and $R(s_0)+(F-B)\tp\cdot\parikh_E(s)<B(t)$ ($\neg \M(s)\step{t}$).
With variables $F\ge\zero$, $B\ge\zero$, and $R(s_0)\ge 0$ a solution
forms a region solving the ESSP. 
\ENDDEF{d.synth}

If the LTS is finite, the (also finite)
linear inequality systems can be solved by standard means, e.g.\ employing 
an ILP- or SMT-solver~\cite{smtinterpol}. If an LTS is not reachable or not deterministic, it cannot be structurally isomorphic
to a reachability graph, i.e.\ no Petri net solving it can be found.
For these reasons, we will assume all LTS to
be finite, reachable, and deterministic in the remainder of this paper.

Fig.~\ref{f.gen1} shows an \lts (left side) and a \RAC Petri net (right side) with a reachability graph
isomorphic to the \lts. All separation problems are solvable. The SSP $(s_4,s_5)$ e.g.\
is solved by the region $r=(R,B,F)$ with $R(s_0)=0$, $F(a)=1=B(d)$, and all other entries being zero.
With $s_0\step{ac}s_4$ and $s_0\step{ade}s_4$ we find 
$R(s_4)=R(s_0)-B(a)+F(a)-B(c)+F(c)=1\neq 0=R(s_0)-B(a)+F(a)-B(d)+F(d)-B(e)+F(e)=R(s_5)$.
The region $r$ also fulfills the cycle conditions $F(b)-B(b)+F(c)-B(c)=0$ and
$F(a)-B(a)+F(d)-B(d)+F(f)-B(f)=0$ since $bc$ and $adf$ are cycles in the \lts.
It corresponds to the place $p_3$.

For the ESSP $(s_{13},a)$ with $s_0\step{adeade}s_{13}$ and $\neg s_{13}\step{a}$ the region
$r=(R,B,F)$ with $R(s_0)=2$, $B(a)=1=F(f)$, and all other entries zero is a solution.
We obtain inequalities
$R(s_{13})=R(s_0)-2B(a)+2F(a)-2B(d)+2F(d)-2B(e)+2F(e)=2-2=0<1=B(a)$ forbidding $a$ at $s_{13}$
and allowing it at $s_0$, $s_4$, and $s_5$ by $R(s_0)=2\ge 1=B(a)$, $R(s_4)=R(s_0)-B(a)+F(a)-B(c)+F(c)=2-1\ge 1=B(a)$,
and $R(s_5)=R(s_0)-B(a)+F(a)-B(d)+F(d)-B(e)+F(e)=2-1\ge 1=B(a)$.
The cycle conditions $F(a)-B(a)+F(d)-B(d)+F(f)-B(f)=0$ and $F(b)-B(b)+F(c)-B(c)=0$ are also fulfilled.
This region $r$ corresponds to the place $p_1$, which prohibits $a$ at $s_{13}$ but not at the three
states where it must occur.

\section{Synthesis by implication and deactivation properties}\label{sect.3}
\newcommand\IP[3]{\ensuremath{\mathrm{I#1}_{#2,#3}}\xspace}
\newcommand\DP[2]{\ensuremath{\mathrm{D}_{#1,#2}}\xspace}
\newcommand\IPS[1]{\ensuremath{\mathrm{I#1}}\xspace}
\newcommand\DPS{\ensuremath{\mathrm{D}}\xspace}

In this section we will show that membership of a net in \WPI or \BRAC
enforces some structural properties in its reachability graph. For pairs of
labels, these properties determine whether the corresponding transitions
in the net must have identical, properly included, or disjoint presets.

\DeclareRobustCommand\rhddisj{\mathrel{\scriptstyle\|\hspace*{0.4mm}}\joinrel\mathrel\rhd}
\DEF{d.impl}{Implication Properties}
For any finite, reachable, deterministic \lts $(S,\Sigma,\to,s_0)$ define the following relations
over $\Sigma$. For two labels $a,b\in\Sigma$, let:
\begin{itemize}
\item $a\equiv b$ $\iff$ $\forall s\in S\colon s\step{a} \iff s\step{b}$,
\item $a\rhd b$ $\iff$ $(\forall s\in S\colon s\step{a} \then s\step{b} \wedge \exists s\in S\colon s\step{b} \wedge \neg s\step{a})$,
\item $a\bowtie b$ $\iff$ ($a\rhd b \vee b\rhd a$), and
\item $a\interleave b$ $\iff$ ($\exists s\in S\colon s\step{a}\wedge\neg s\step{b} \wedge \exists s\in S\colon s\step{b}\wedge\neg s\step{a}$).
\end{itemize}
\ENDDEF{d.impl}

Take a look at the \lts on the left side of Fig.~\ref{f.gen1}. The labels $e$ and $f$ both occur at
$s_2$, $s_{10}$, and $s_{12}$ but nowhere else, thus we get $e\equiv f$. At $s_0$, $s_4$, and $s_5$
both $a$ and $b$ are allowed, but $b$ is also present at $s_9$, $s_{11}$, and $s_{13}$. We conclude
$a\rhd b$ and thus $a\bowtie b$. Similarly, $c$ and $d$ are possible at $s_1$, $s_6$, and $s_7$, and further $c$'s occur
at $s_3$, $s_8$, and $s_{14}$, so $d\rhd c$ and $c\bowtie d$ hold. All other pairs of labels are related via $\interleave$.
Looking at the synthesised Petri net in the figure, we find that $e\equiv f$ suggests identical
presets for $e$ and $f$, while for $a\rhd b$ and $d\rhd c$ we find proper inclusion of presets.
Pairs of labels in the $\interleave$-relation have disjoint presets in the Petri net. 

The three relations $\equiv$, $\bowtie$, and $\interleave$ obviously partition the set of pairs of labels
for any \lts.

\KOR{c.impl}{Implication Properties are mutually exclusive}
For any finite, reachable, deterministic LTS $(S,\Sigma,\to,s_0)$ and two
labels $a,b\in\Sigma$, exactly one of the three properties $a\equiv b$,
$a\bowtie b$, and $a\interleave b$ holds.
Also, $a\rhd b$ and $b\rhd a$ are mutually exclusive, i.e.\ exactly one of them is true if $a\bowtie b$.
\ENDKOR{c.impl}

To detect whether two transitions must have a common place in their preset, the following
property is useful.

\DEF{d.dp}{Deactivation Property}
For any finite, reachable, deterministic LTS $(S,\Sigma,\to,s_0)$ and two
labels $a,b\in\Sigma$, we define the relation $\merge$ by
$a\merge b$ $\iff$ $\exists s,s'\in S\colon (s\step{a}\wedge s\step{b}s' \wedge\neg s'\step{a})
 \vee (s\step{b}\wedge s\step{a}s' \wedge\neg s'\step{b})$.
\ENDDEF{d.dp}

Assume now a given \lts $TS=(S,\Sigma,\to,s_0)$ which is finite, reachable, and deterministic
as well as two labels $a,b\in\Sigma$.
We check each combination of one of $a\equiv b$, $a\bowtie b$, and $a\interleave b$
with either $a\merge b$ or $\neg a\merge b$ to find out how far this determines the structure
of a possible synthesized net in the class \WPI or \BRAC. Overall, we have six cases.

{\bf Case 1:} $a\interleave b \wedge a\merge b$.\\
From $a\merge b$ we conclude that the presets of transitions $a$ and $b$ in a \WPI or \BRAC net $N=(P,T,W,M_0)$
that solves $TS$ are not disjoint, as one can deactivate the other. This contradicts
$a\interleave b$, telling us that neither $W(\cdot,a)\le W(\cdot,b)$ nor $W(\cdot,b)\le W(\cdot,a)$
hold, i.e.\ the presets of $a$ and $b$ must be disjoint.
Synthesis must fail in this case, there exists no \WPI or \BRAC net solving $TS$.
Note that this case may occur in reachability graphs of \ACC nets.
Then, the presets of $a$ and $b$ would necessarily overlap, but not be contained in one another.

{\bf Case 2:} $a\interleave b \wedge \neg a\merge b$.\\
Like argued in the previous case, $a\interleave b$ dictates that transitions $a$ and $b$ must have disjoint presets
in any \WPI net solving $TS$.

{\bf Case 3:} $a\equiv b \wedge a\merge b$.\\
As one of the labels can deactivate the other in the LTS, the transitions $a$ and $b$ in a
Petri net $N=(P,T,W,M_0)$ solving $TS$ must have a common place $p$ with $W(p,a)>0$ and $W(p,b)>0$. So, the
presets of $a$ and $b$ are either identical or there is a proper inclusion. Assume the preset
of $b$ is properly included in that of $a$, i.e.\ we have an asymmetric choice.
We construct a new Petri net $N'=(P,T,W',M_0)$.
For every place $q$ with $W(q,a)>W(q,b)$ we set $W'(q,b)=W(q,a)$ and $W'(b,q)=W(b,q)+W(q,a)-W(q,b)$.
For all other combinations of $x,y\in P\cup T$ let $W'(x,y)=W(x,y)$. 
If $N$ is \WPI, so is $N'$ since the transition $a$'s preset compares correctly to all other transitions
and $b$ gets this same preset. If $N$ is \BRAC, so is $N'$, as the asymmetric choice between $a$ and $b$
is removed in $N'$ compared to $N$. In the definition of \BRAC, this means $b$ moves from the block $T_2$
of transitions to $T_1$, but the condition remains true.
Firing any transition changes the marking
in the same way in both nets, $N$ and $N'$, as $W(x,y)-W(y,x)=W'(x,y)-W'(y,x)$ for all $x,y\in P\cup T$.
All transitions except $b$ are either enabled both or disabled both in $N$ and $N'$ under
the same marking. Under any marking, the transition $b$ is enabled in $N$ if it is enabled in $N'$.
Now assume $b$ is enabled in $N$ and for some state $s\in S$ with $s\step{b}$ 
and the marking $M(s)$ 
corresponding to $s$ in $N$ we have $M(s)\ge W(\cdot,b)$.
Then, due to $a\equiv b$, we get $s\step{a}$ and $M(s)\ge W(\cdot,a)=W'(\cdot,b)$.
Thus, $b$ is enabled under the same markings in $N$ and $N'$ and the reachability graphs of $N$
and $N'$ must be isomorphic. So, if $N$ solves $TS$, also $N'$ solves $TS$ with $W'(q,a)=W'(q,b)$.
By symmetry, the same argument holds for reversed roles of $a$ and $b$, so synthesis is always possible
in such a way that $a$ and $b$ have an identical preset (considering arc weights).

{\bf Case 4:} $a\equiv b \wedge \neg a\merge b$.\\
If no state $s\in S$ with $s\step{a}$ and $s\step{b}$ exists, $a$ and $b$ do not occur at all
in $TS$ and can be eliminated from the alphabet $\Sigma$. Now let $s\in S$ be a state with
$s\step{a}s'$ and $s\step{b}$. As $a$ does not deactivate $b$, $s'\step{b}$ and thus also $s'\step{a}$ holds.
If $a$ is activated at $s$, it is activated at its successor, and by induction at all states
reachable from $s$ via $a$'s only. 
As the \lts is finite, there must be some $a$-cycle with
$s\step{a^m}s'\step{a^n}s'$ for some $m,n\in\N$ with $n\ge 1$. We conclude
$M(s')=M(s')+n\cdot(W(a,\cdot)-W(\cdot,a))$ by the firing rule. We obtain $W(\cdot,a)=W(a,\cdot)$
and $a$ leaves the marking unchanged, i.e.\ $M(s')\step{a}M(s')$ as well as $M(s)\step{a}M(s)$. 
As $s$ is the unique state with marking $M(s)$, we find $s\step{a}s$ in the LTS. 
Therefore, $a$ (and by symmetry also $b$) forms self-loops wherever in the \lts it occurs.

Assume now a \WPI or \BRAC net $N=(P,T,W,M_0)$ solving $TS$ such that the presets of $a$ and $b$
are properly contained in one another or disjoint. We construct a new net
$N'=(P,T,W',M_0)$ with $W'(\cdot,a)=W'(a,\cdot)=W(\cdot,b)$ and for $t\in T\backslash\{a\}$ with
$W'(\cdot,t)=W(\cdot,t)$ and $W'(t,\cdot)=W(t,\cdot)$. As $a$ now has the same preset as $b$,
$N'$ is in \WPI. If $N$ is in \BRAC, $b$ may be involved in an asymmetric choice (i.e.\ for some places, $b$
appears in one of the blocks $T_1$ or $T_2$ of transitions as per definition of \BRAC). Since $a$
gets the same preset as $b$ in $N'$, it becomes a member in the same block as $b$, and the
defining conditions for \BRAC also hold for $N'$.\footnote{This does not work for synthesis 
of \RAC-nets. If ${}^\bullet a\cap{}^\bullet b=\emptyset$ in $N$, $b$ could already be involved
in an asymmetric choice via some $c$ with ${}^\bullet c\subseteq{}^\bullet b$. In $N'$, we would
get ${}^\bullet c\subseteq{}^\bullet a={}^\bullet b$, violating the \RAC condition.} 
If $b$ is part of a free-choice block of
transitions (as in the left of Fig.~\ref{f.brac}), $a$ becomes a member in this block, too,
again as it has now the same preset as $b$. In both cases, $N'$ is in \BRAC.
Since $a$ and $b$ occur at the
same reachable states $s$ and markings $M(s)$ of $N$ and $N'$, $a$ is still activated in $N'$
exactly when $b$ is. So, if the \lts $TS$ is synthesisable, we find a solution where $a$ and $b$
have identical presets.

{\bf Case 5:} $a\bowtie b \wedge a\merge b$.\\
As one of $a$ and $b$ can deactivate the other due to $a\merge b$, their presets in a \WPI net
$N=(P,T,W,M_0)$ solving $TS$ cannot be disjoint. As $a\bowtie b$ proves the existence of a state
$s\in S$ with $s\step{a}$ and $\neg s\step{b}$ (or the other way around), the presets of
$a$ and $b$ cannot be identical. Therefore, $W(\cdot,a)\ge W(\cdot,b)$ or $W(\cdot,a)\le W(\cdot,b)$,
depending on which label occurs at $s$ (i.e.\ whether $a\rhd b$ or $b\rhd a$ holds).
In both cases, we have proper containment between the presets and know in which direction.

{\bf Case 6:} $a\bowtie b \wedge \neg a\merge b$.\\
W.l.o.g. assume $b\rhd a$ ($a\rhd b$ is handled symmetrically).
In a \WPI/\BRAC net $N=(P,T,W,M_0)$ solving $TS$,
$W(\cdot,a)\neq W(\cdot,b)$. So, either the presets are disjoint or
$W(\cdot,a)\lneqq W(\cdot,b)$. Let $s\in S$ be a state with
$s\step{a}$ and $s\step{b}$ (otherwise $b$ would not occur in $TS$ at all).
As we have already seen in case~4, if $a$'s form a cycle in the \lts, all $a$'s are automatically self-loops.
If $a$ is not a self-loop at $s$, there is a maximal $k\in\N$ with
$s\step{a^k}s'$ (due to finiteness of $TS$).
If $W(\cdot,a)\le W(\cdot,b)$ held, we would find that with
$\neg s'\step{a}$ also $\neg s'\step{b}$ was true, so somewhere
in the walk $s\step{a^k}s'$ the label $b$ got deactivated by $a$,
contradicting $\neg a\merge b$. Thus, when $a$ is not a self-loop,
we find that $a$ and $b$ have disjoint presets in $N$.

If $a$ forms a self-loop at $s$, we conclude $M(s)\step{a}M(s)$
in $N$ for the marking $M(s)$ corresponding to $s$. Therefore,
$W(\cdot,a)=W(a,\cdot)$. The presets of $a$ and $b$ may be disjoint or $W(\cdot,a)\lneqq W(\cdot,b)$.
With the help of Fig.~\ref{f.case6}, we can show that both situations may occur and that we
may not be able to choose whether we want disjointness or proper containment.

\tikzset{loop/.style={to path={.. controls +(315:1) and +(45:1) .. (\tikztotarget) \tikztonodes}}}
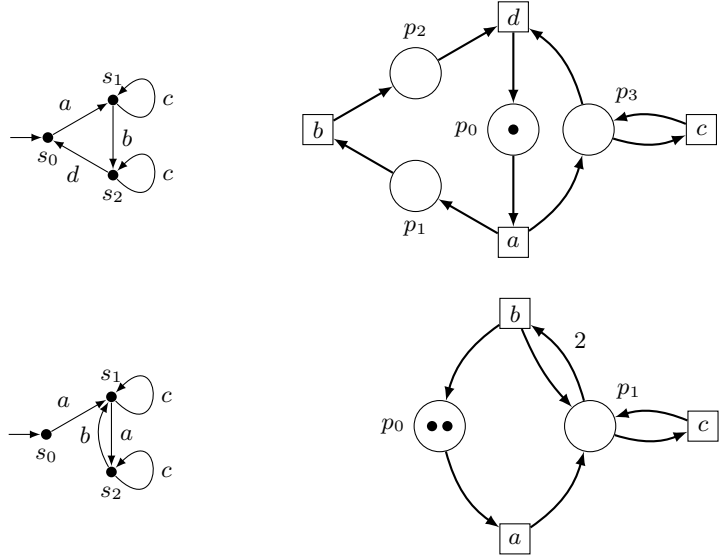
\begin{figure}[t]
\centering
\begin{tikzpicture}[LTS]
\node[state,label=below:$s_0$] (s0) {};
\path[edge] (s0) ++(-0.5,0) edge (s0);
\path(s0) +(30:1cm) node[state,label=above:$s_1$] (s1) {};
\path(s0) +(-30:1cm) node[state,label=below:$s_2$] (s2) {};
\draw[edge] (s0) edge node[auto]{$a$} (s1);
\draw[edge] (s1) edge node[auto]{$b$} (s2);
\draw[edge] (s2) edge node[auto,xshift=1mm]{$d$} (s0);
\draw[edge] (s1) edge[loop] node[auto,swap]{$c$} (s1);
\draw[edge] (s2) edge[loop] node[auto,swap]{$c$} (s2);
\path(s0) to (0,-1.6);
\end{tikzpicture}\hspace*{1.5cm}
\begin{tikzpicture}
\draw(0,0) node[transition] (b) {$b$};
\draw(0,0) +(30:1.5) node[place,label=above:$p_2$] (p2) {};
\draw(0,0) +(-30:1.5cm) node[place,label=below:$p_1$] (p1) {};
\draw(0,0) +(-30:3cm) node[transition] (a) {$a$};
\draw(0,0) +(30:3cm) node[transition] (d) {$d$};
\draw(0,1.5) +(-30:3cm) node[place,tokens=1,label=left:$p_0$] (p0) {};
\draw(1,1.5) +(-30:3cm) node[place,label=above right:$p_3$] (p3) {};
\draw(2.5,1.5) +(-30:3cm) node[transition] (c) {$c$};
\draw(p1) edge[-latex,thick] (b);
\draw(b) edge[-latex,thick] (p2);
\draw(p2) edge[-latex,thick] (d);
\draw(d) edge[-latex,thick] (p0);
\draw(p0) edge[-latex,thick] (a);
\draw(a) edge[-latex,thick] (p1);
\draw(a) edge[-latex,thick,bend right=20] (p3);
\draw(p3) edge[-latex,thick,bend right=20] (d);
\draw(c) edge[-latex,thick,bend right=20] (p3);
\draw(p3) edge[-latex,thick,bend right=20] (c);
\end{tikzpicture}\\[5mm]
\begin{tikzpicture}[LTS]
\node[state,label=below:$s_0$] (s0) {};
\path[edge] (s0) ++(-0.5,0) edge (s0);
\path(s0) +(30:1cm) node[state,label=above:$s_1$] (s1) {};
\path(s0) +(-30:1cm) node[state,label=below:$s_2$] (s2) {};
\draw[edge] (s0) edge node[auto]{$a$} (s1);
\draw[edge] (s1) edge node[auto]{$a$} (s2);
\draw[edge] (s2) edge[bend left=30] node[auto]{$b$} (s1);
\draw[edge] (s1) edge[loop] node[auto,swap]{$c$} (s1);
\draw[edge] (s2) edge[loop] node[auto,swap]{$c$} (s2);
\path(s0) to (0,-1.6);
\end{tikzpicture}\hspace*{1.5cm}
\begin{tikzpicture}
\path(-0.25,0);
\draw(0,0) +(-30:3cm) node[transition] (a) {$a$};
\draw(0,0) +(30:3cm) node[transition] (b) {$b$};
\draw(0,1.5) ++(-30:3cm) ++(-1,0) node[place,tokens=2,label=left:$p_0$] (p0) {};
\draw(1,1.5) +(-30:3cm) node[place,label=above right:$p_1$] (p1) {};
\draw(2.5,1.5) +(-30:3cm) node[transition] (c) {$c$};
\draw(p1) edge[-latex,thick,bend right=20] node[auto,swap]{$2$} (b);
\draw(b) edge[-latex,thick,bend right=10] (p1);
\draw(b) edge[-latex,thick,bend right=20] (p0);
\draw(p0) edge[-latex,thick,bend right=20] (a);
\draw(a) edge[-latex,thick,bend right=20] (p1);
\draw(c) edge[-latex,thick,bend right=20] (p1);
\draw(p1) edge[-latex,thick,bend right=20] (c);
\end{tikzpicture}
\caption{Left: Two \lts (left) with $b\rhd c$ and $\neg b\merge c$. Right: \WPI Petri net solutions 
for the \lts on the left. The upper net is also in (B)\RAC}
\label{f.case6}
\end{figure}

In the upper part of Fig.~\ref{f.case6} we see an \lts on the left with $b\rhd c$ and $\neg b\merge c$.
The \lts is synthesisable to a \WPI net shown on the right. In this net, the presets of $b$ and $c$ 
are disjoint. There is no solution where the preset of $c$ is properly contained in that of $b$.
Note that $b\interleave d$, $\neg b\merge d$, $d\rhd c$, and $d \merge c$ hold.
By case 2, $b$ and $d$ have disjoint presets, and by case 5, the preset of $c$ is properly contained
in that of $d$, so $b$ and $c$ must have disjoint presets.

In the lower part of Fig.~\ref{f.case6} we have again an \lts with $b\rhd c$ and $\neg b\merge c$,
with a \WPI net solving it on the right side. Here, the preset of $c$ is properly contained in that
of $b$. We see that $a\interleave c$ and $\neg a\merge c$ imply disjoint presets by case~2. If $b$ and
$c$ also had disjoint presets, only $c$ could remove tokens from its own preset. But as $c$ is
a self-loop, this does not happen either. As $a$ activates $c$ by its first occurrence, it must
put a token in $c$'s preset and then $s_1\step{ab}s_1$ would increase this number of
tokens, contradicting the fact that it is a cycle and the Petri net to be constructed must always be at the
same marking when reaching state $s_1$. Thus, no solution with disjoint presets for $b$ and $c$
exists.

That we cannot simply decide for disjointness or proper containment when self-loops are involved
in case~6 poses a problem for synthesis, which we bypass for the
remainder of this section by disallowing self-loops in the \lts.
We might argue that such self-loops are not too interesting as they typically denote idle actions anyway, 
not changing the system's state. Note that for a solvable \lts all edges with 
the same label are synthesised to a single transition, so either all or none of them
must be a self-loop.
For \lts without self-loops we find:

\SATZ{l.cases1}{Label relations in the self-loop free case}
Let $TS=(S,\Sigma,\to,s_0))$ be an \lts without self-loops. For every pair of labels
$a,b\in\Sigma$ we can determine whether the transitions $a$ and $b$ in a \WPI or \BRAC
net solving $TS$ have disjoint, identical, or properly contained presets, i.e. which
of the four relations ${}^\bullet a\cap {}^\bullet b=\emptyset$, ${}^\bullet a ={}^\bullet b$,
${}^\bullet a \subsetneqq {}^\bullet b$, or ${}^\bullet a \supsetneqq {}^\bullet b$ holds.
\ENDSATZ
\BEW
We have seen above that any pair $(a,b)$ falls under one of six cases.
Case~1 immediately renders the \lts non-synthesisable and case~4 can only occur
if self-loops are present. 
Case~2 and~6 (without self-loops) imply disjoint presets and case~3 determines them to be identical.
In the remaining case~5, one of the presets is contained in the other and
$a\rhd b$ implies ${}^\bullet a \supsetneqq {}^\bullet b$ while
$b\rhd a$ implies ${}^\bullet a \subsetneqq {}^\bullet b$.
\ENDBEW{l.cases1}

\subsection{Solving separation problems for \WPI synthesis}

We can now formulate inequality systems for separation problems that incorporate the above relations
between labels. Since there is no limit on the number of places in the preset of a transition for
\WPI nets, we can tackle each separation problem with its own inequality system and find a region
solving it.
For an ESSP $(s,a)$ in an \lts $TS=(S,\Sigma,\mathord{\to},s_0)$ we then obtain

\begin{eqnarray*}
                              && R(s_0)+(F-B)\tp\cdot\parikh_E(s) < B(a) \\
\forall s'\step{t}s''\in\mathord{\to}\colon   && R(s_0)+(F-B)\tp\cdot\parikh_E(s')\ge B(t) \\
\forall \gamma\in\Gamma\colon && (F-B)\tp\cdot\gamma = 0 \\
\forall b\in T\colon a\equiv b\then && B(b)=B(a) \\
\forall b\in T\colon a\rhd b\wedge a\merge b\then && B(b)\le B(a) \\
\forall b\in T\colon b\rhd a\wedge a\merge b\then && B(a)\le B(b) \\
\forall b\in T\colon a\bowtie b \wedge \neg a\merge b\then && B(b)=0 \\
\forall b\in T\colon a\interleave b \wedge \neg a\merge b\then && B(b)=0 
\end{eqnarray*}
The last five lines stand for the cases~3, 5, 5, 6, and~2, in this order.
For an SSP $(s,s')$ the first line changes to
\begin{eqnarray*}
(F-B)\tp\cdot(\parikh_E(s)-\parikh_E(s')) & \neq & 0. \\
\end{eqnarray*}
Since the label $a$ is not part of the problem instance for the SSP $(s,s')$,
we must construct inequality systems for each label $a\in T$. If at least one of them
is solvable, the SSP is also solved, and we can construct a place from one of the solutions.

\subsection{Solving separation problems for \BRAC synthesis}

When targetting a \BRAC net, the situation is more complicated. First, we must forbid
arc weights by adding
$\forall t\in T\colon B(t)\le 1 \wedge F(t)\le 1$ to every inequality system we create.
The constant terms $1$ in these inequalities make solving these inequality systems \NP-complete.
Then, we must ensure that transitions involved in an asymmetric choice have only one
or two places in their preset. So, if case~5 occurs with $a\bowtie b \wedge a\merge b$
for some labels $a$ and $b$, we may construct only 
two inequality systems for all separation problems concerning $a$ and $b$ together.
We start by solving ESSPs and take a look at SSPs afterwards.


Assume, targetting at \BRAC, we have an asymmetric choice for two labels $a,b\in T$ with $a\rhd b$ and $a\merge b$
(since case~5 is the only case with properly included presets).
All ESSPs $(s,b)$ with $\neg s\step{b}$ must be solved by the same region, i.e.\ a single place.
If $b$ can occur exactly $k$ times consecutively at a state $s$,
$s\step{b^k}$ and $\neg s\step{b^{k+1}}$, we have $R(s)=k$, $B(b)=1$, and $F(b)=0$.\footnote{Otherwise
$b$ would be a self-loop or the solving net would be unbounded.}
Furthermore, $B(a)=1$ (by case~5) and all $t\in T$ with either $a\equiv t$ or $b\equiv t$
($t$'s with presets identical to that of either $a$ or $b$ by case~3 or~4) also yield $B(t)=1$.
For all other transitions $t'\in T$ we get $B(t')=0$ (by the remaining cases~2 or~6).
Because $a$ will have two places in its preset, $F(a)=1$ does not mean that $a$ is a self-loop,
i.e.\ the value of $F(a)$ is not predetermined. The same holds for $t\in T$ with $a\equiv t$.
We solve the inequality system from above, modified by the known, fixed values and one copy of
the first line, $R(s_0)+(F-B)\tp\cdot\parikh_E(s) < B(b)$, for each ESSP $(s,b)$ with $\neg s\step{b}$,
to obtain the region/place in the preset of $b$. This also solves all ESSPs for $t\in T$ with $b\equiv t$,
as these labels share $B(t)=1$ and $F(t)=0$ as well as the number of consecutive activations at any state $s$, 
i.e. the value $R(s)$, with $b$.

A second region must solve all ESSPs $(s,a)$ with $\neg s\step{a}$ but $s\step{b}$ at the same time.%
\footnote{ESSPs $(s,a)$ with $\neg s\step{b}$ are already solved by the first region since ${}^\bullet b\subsetneqq {}^\bullet a$.}
We add the first line of our inequality system,
$R(s_0)+(F-B)\tp\cdot\parikh_E(s) < B(a)$, for each state $s$ with such an ESSP $(s,a)$.
Labels $t$ with $a\equiv t$ share this line with $a$ as $B(a)=B(t)$, so all ESSPs $(s,t)$ are tackled
at the same time automatically.\footnote{The values $F(a)$ and $F(t)$ for $a\equiv t$, which may actually
differ, are mostly determined via $F\cdot\parikh_E(s')$ for different $s'$ in the first and second line of our generic 
inequality system, but also restricted by the cycles of $\Gamma$ in the third line.}
A solution, if it exists, forms the second place in the preset of $a$.%
\footnote{If any constructed inequality system is unsolvable, so is the corresponding ESSP, 
and the \lts is not synthesisable.}

All labels not involved in asymmetric choices form blocks with presets disjoint to each other (and from the asymmetric
choices), while all labels inside one block have identical presets. ESSPs for labels in these blocks
can be handled using the generic inequality system from above. Each ESSP may necessitate its own region and
thus its own place.

SSPs $(s,s')$ are best solved apart from asymmetric choice blocks first. We simply replace
the two lines for properly included presets in our above generic inequality system for SSPs/\BRAC by
\begin{eqnarray*}
\forall b\in T\colon a\rhd b\wedge a\merge b\then && B(a)=0 \wedge B(b)=0 \\
\forall b\in T\colon b\rhd a\wedge a\merge b\then && B(a)=0 \wedge B(b)=0 
\end{eqnarray*}
If such a system has a solution, we can convert the found region to a place in one
of the free-choice blocks in our \BRAC net. Otherwise, the SSP $(s,s')$ must be
solved by an asymmetric choice block. If no region constructed so far solves this SSP, one of the regions we have
generated for an asymmetric choice block above is not sufficient. Since there may be more than one unsolved SSP,
we must solve those inequality systems again, and for each unsolved SSP $(s,s')$
we must either add the line $(F-B)\tp\cdot(\parikh_E(s)-\parikh_E(s')) \neq 0$ or
leave it out (if the SSP will be solved via the other of the two regions or even
in a different asymmetric choice block), in all combinations. Clearly, this becomes exponential. Still,
guessing which unsolved SSP will be solved by which asymmetric choice block and which of two regions
is in \NP, so the whole problem could be \NP-complete in the worst case. 

\KOR{k.sfsynth}{Synthesis to self-loop free \WPI or \BRAC nets}
A \WPI or \BRAC net can be synthesised from an \lts by solving all modified separation problems (as shown above)
and converting solving regions for the inequality systems into places canonically.
While solving for \WPI nets has a polynomial time complexity, the extra condition for \BRAC nets
involving constant terms in the inequality systems only allows us to deduce membership in \NP.
\ENDKOR{k.sfsynth}

So, we have two problems that may prevent a polynomial runtime for \BRAC synthesis: 
the limit on arc weights and the combinatorial problem of SSPs and asymmetric choice blocks.

We can attempt to solve each one of these problems (but not both at the same time) by extending the class \BRAC.
To allow solving the SSPs in polynomial time, we can extend the class such that an asymmetric
choice block $T'$ of transitions can have more than two places in its preset. Each of the two 
subblocks of transitions, $T'=T_1\cup T_2$ with $T_1\cap T_2=\emptyset$, must have a common preset, i.e. 
$\forall i\in\{1,2\}\forall t,t'\in T_i\colon {}^\bullet t={}^\bullet t'$.
Between blocks, proper containment of presets holds, i.e.
$\forall t_1\in T_1, t_2\in T_2\colon {}^\bullet t_2\subsetneqq{}^\bullet t_1$.
Under the condition that all arc weights (in the presets) are equal to one, we can formulate
two inequality systems for each (unsolved) SSP, one for the sought region/place in the preset of $T_1$,
the other for $T_2$ only. If one is solved, we can add a place accordingly to the constructed net.
If no solution exists, the SSPs must be handled in some other asymmetric choice block.

We can also try to remove the condition $\forall t\in T\colon F(t)\le 1\wedge B(t)\le 1$ to make our
inequality systems solvable in polynomial time. But this will require us to define weighted
asymmetric choices that also respect the \WPI condition. In an asymmetric choice block
$T'=T_1\cup T_2$ with two places $p$, $p'$ with $p^\bullet=T_1$ and $p'{}^\bullet=T'$ we must
demand $W(p,t_1)\le W(p',t_1)\ge W(p',t_2)$ for all $t_1\in T_1$ and $t_2\in T_2$.
The first condition can only be formulated if we merge the two inequality systems for an
asymmetric choice block into one (with still two separate sets of variables). Then we can add
the above comparisons. 

Now consider what happens if we make both extensions to \BRAC at the same time.
We might have two (or more) different, so far unsolved SSPs that can be addressed
in the same asymmetric choice block. With arc weights limited to one, we just had to check
if an SSP could be solved with one of the two predetermined presets, but possibly different 
postsets for the transitions.
With arbitrary arc weights, the solutions might now require different arc weights
in the preset of the asymmetric choice block.
Since places can still only have one of two different postsets in an asymmetric choice,
we must decide which SSPs need to be solved by {\it this} asymmetric choice block. 
Between all asymmetric choices and all SSPs to solve, we arrive again at a combinatorial problem.

We make the following conjecture.

\CON{c.2}{Extensions of \BRAC}\\
For every extension of the class \BRAC still allowing only weighted asymmetric choices,
the synthesis is \NP-complete.
\ENDCON{c.2}

\section{Handling self-loops in the \lts}\label{sect.4}

\tikzstyle{myRightArrowBaseline}=[baseline=-0.6ex]
\tikzstyle{myFullArrow}=[-stealth]
\tikzstyle{myBackArrow}=[stealth-]
\tikzstyle{myLine}=[-]
\tikzstyle{myDashedArrow}=[myFullArrow, dash pattern=on 1.5pt off 1.5pt]
\tikzstyle{myBackDashedArrow}=[myBackArrow, dash pattern=on 1.5pt off 1.5pt]
\tikzstyle{myNone}=[]
\newcommand{\myArrow}[1]{\mathbin{\tikz [myRightArrowBaseline]{
        \draw [#1] (0,0) -- (1.2em,0);
}}}
\newcommand{\myNoArrow}{\mathbin{\tikz [myRightArrowBaseline]{
        \path (0,0) -- (1.2em,0);
}}}
\newcommand{\disjorincl}{\myArrow{myDashedArrow}}
\newcommand{\included}{\myArrow{-stealth}}
\newcommand{\backdisjorincl}{\myArrow{myBackDashedArrow}}
\newcommand{\backincluded}{\myArrow{myBackArrow}}
\newcommand{\equivalent}{\myArrow{myLine}}
\newcommand{\disjoint}{\myNoArrow}

\DEF{d.grl}{Graphical preset relations}
For a finite, reachable, deterministic \lts $TS=(S,\Sigma,\to,s_0)$ with two
labels $a,b\in\Sigma$, the six cases in the previous section enforce six
possible preset relations for $a$ and $b$ in a theoretical, synthesised
\WPI net $N=(P,T,W,M_0)$. We denote by
\begin{itemize}
\item $a\equivalent b$ that $W(\cdot,a)=W(\cdot,b)$ via case~3 or~4 (equivalence),
\item $a\disjoint b$ that ${}^\bullet a\cap{}^\bullet b=\emptyset$ via case~2 or~6 (disjointness),
\item $a\included b$ that $W(\cdot,a)\lneqq W(\cdot,b)$ via case~5 (proper inclusion),
\item $a\backincluded b$ that $W(\cdot,a)\gneqq W(\cdot,b)$ via case~5 (proper inclusion),
\item $a\disjorincl b$ that $a$ is a self-loop in $TS$ with either ${}^\bullet a\cap{}^\bullet b=\emptyset$ or $W(\cdot,a)\lneqq W(\cdot,b)$ via case~6 (disjoint or properly included), or
\item $a\backdisjorincl b$ that $b$ is a self-loop in $TS$ with either ${}^\bullet a\cap{}^\bullet b=\emptyset$ or $W(\cdot,a)\gneqq W(\cdot,b)$ via case~6 (disjoint or properly included).
\end{itemize}
Each pair $(a,b)$ is in exactly one of these relations unless case~1 disallows
the synthesis of the \lts at all.
\ENDDEF{d.grl}

We will now study the preset relations between more than two labels. Consider e.g.\
two labels $a$ and $b$ with $a\equivalent b$ and a third label $c$. Intuitively,
we would assume that the relations between $a$ and $c$ and between $b$ and $c$
are the same, as $a$ and $b$ are somehow equivalent. Let us check that.

\LEM{l.equiv}{Adapting equivalence}
Let $TS=(S,\Sigma,\to,s_0)$ be a finite, reachable, deterministic \lts.
For any three labels $a,b,c\in\Sigma$ with $a\equivalent b$, we can
make the following conclusions:
\begin{itemize}
\item $a\equivalent c \then b\equivalent c$,
\item $a\disjoint c \then (b\disjoint c \vee b\disjorincl c \vee b\backdisjorincl c)$,
\item $a\included c \then (b\included c \vee b\disjorincl c)$,
\item $a\backincluded c \then (b\backincluded c \vee b\backdisjorincl c)$,
\item $a\disjorincl c \then (b\disjorincl c \vee b\included c \vee b\disjoint c)$,
\item $a\backdisjorincl c \then (b\backdisjorincl c \vee b\backincluded c \vee b\disjoint c)$,
\end{itemize}
We can strengthen the relations by deducing for any edge $b\disjorincl c$ either that
$b\included c$ holds or that $b\disjoint c$ holds, similarly for reverse edges.
Then, for any of the six above relations $R$ holds $aRc \then bRc$ and
$\equivalent$ becomes an equivalence respecting the other five relations.
\ENDLEM
\BEW
All points follow directly from $W(\cdot,a)=W(\cdot,b)$ and the preset relation
between $a$ and $c$. All relations that allow this preset comparison are then
possible between $b$ and $c$.

In the cases $a\disjoint c$, $a\included c$, and $a\backincluded c$, we know
the exact relation, ${}^\bullet a\cap{}^\bullet c=\emptyset$ or
$W(\cdot,a)\lneqq W(\cdot,c)$ or $W(\cdot,a)\gneqq W(\cdot,c)$, respectively.
We may then conclude that $b$ must be in that same relation to $c$, so
we may strengthen $b\disjorincl c$ to either $b\included c$ or $b\disjoint c$,
and $b\backdisjorincl c$ to either $b\backincluded c$ or $b\disjoint c$.
This removes all dashed edges from the first four points.

We can also remove the relations $b\included c$, $b\backincluded c$, and
$b\disjoint c$ from the last two points by symmetry (swapping the roles of $a$ and $b$).
As a total effect, $a$ and $b$ then are in the same relation to $c$.
\ENDBEW{l.equiv}

This means after checking for inconsistencies (case~1) and strengthening
$\disjorincl$ and $\backdisjorincl$ wherever possible, we can deal with
one representative from each equivalence class of $\equivalent$ and forget
about the latter relation at all.

Note that edges $a\disjorincl b$ that were strengthend to $a\included b$ do not share
the deactivation property with original edges $a\included b$. For the former we have
$\neg a\merge b$ while for the latter $a\merge b$ holds. When strengthening the
equivalence $\equivalent$ in this way as shown in Lemma~\ref{l.equiv}, at least
one member in an equivalence class of labels has an original edge $\included$,
and we should choose this label as representative for the class. The deactivation
property will allow us to draw additional conclusions.

Ignoring equivalence, two labels can be related in five ways. If we consider
our problem case $a\disjorincl b$, the two labels can be related to a third
label $c$ in $5^2=25$ ways. Figure~\ref{f.25} shows all possibilities.
Some of these lead directly to a contradiction, shown by the symbol $\times$
inside the triangle. Contradictions happen either because the arrows form
a cycle\footnote{Impossible as $x\rhd y$ in case~5 and~6 means that the \lts contains more
edges with label $y$ than with label $x$} as in 8), 10), 18), and 20),
or because self-loops $a$ and $c$ need to deactivate each other, as in 22) and 23).
In 12) proper inclusion follows from transitivity while in 13) disjointness for $a$ and $b$
contradicts the \WPI property. For 3) and 6) proper inclusion for $a$ and $b$ contradicts 
transitivity and in 2) again the \WPI property comes into play.
Finally, in 17) $a$ cannot deactivate $c$ as it is a self-loop, thus $c$ can deactivate $a$.
If $a\disjorincl b$ means proper inclusion, the preset of $a$ will be contained in both
of $b$ and $c$, i.e. the edge $b\disjorincl c$ also means proper inclusion. As $c$ can
deactivate $a$, it can also deactivate $b$ then, contradicting $b\disjorincl c$ and case~6.

\begin{figure}[t]
\centering
\begin{tikzpicture}
\draw(0,8) node(a11) {$a$};
\draw(a11) +(90:1cm) node(b11) {$b$};
\draw(a11) +(30:1cm) node(c11) {$c$};
\draw[-latex,dashed] (a11) -- (b11);
\draw(a11) +(60:5.5mm) node {?};
\draw(a11) +(135:6.5mm) node {$1)$};

\draw(2.4,8) node(a21) {$a$};
\draw(a21) +(90:1cm) node(b21) {$b$};
\draw(a21) +(30:1cm) node(c21) {$c$};
\draw[-latex,dashed] (a21) -- (b21);
\draw[-latex] (a21) -- (c21);
\draw(a21) +(60:5.5mm) node {$\interleave$};
\draw(a21) +(135:6.5mm) node {$2)$};

\draw(4.8,8) node(a31) {$a$};
\draw(a31) +(90:1cm) node(b31) {$b$};
\draw(a31) +(30:1cm) node(c31) {$c$};
\draw[-latex,dashed] (a31) -- (b31);
\draw[-latex] (c31) -- (a31);
\draw(a31) +(60:5.5mm) node {$\interleave$};
\draw(a31) +(135:6.5mm) node {$3)$};

\draw(7.2,8) node(a41) {$a$};
\draw(a41) +(90:1cm) node(b41) {$b$};
\draw(a41) +(30:1cm) node(c41) {$c$};
\draw[-latex,dashed] (a41) -- (b41);
\draw[-latex,dashed] (a41) -- (c41);
\draw(a41) +(60:5.5mm) node {?};
\draw(a41) +(135:6.5mm) node {$4)$};

\draw(9.6,8) node(a51) {$a$};
\draw(a51) +(90:1cm) node(b51) {$b$};
\draw(a51) +(30:1cm) node(c51) {$c$};
\draw[-latex,dashed] (a51) -- (b51);
\draw[-latex,dashed] (c51) -- (a51);
\draw(a51) +(60:5.5mm) node {?};
\draw(a51) +(135:6.5mm) node {$5)$};

\draw(0,6) node(a12) {$a$};
\draw(a12) +(90:1cm) node(b12) {$b$};
\draw(a12) +(30:1cm) node(c12) {$c$};
\draw[-latex,dashed] (a12) -- (b12);
\draw[-latex] (b12) -- (c12);
\draw(a12) +(60:5.5mm) node {$\interleave$};
\draw(a12) +(135:6.5mm) node {$6)$};

\draw(2.4,6) node(a22) {$a$};
\draw(a22) +(90:1cm) node(b22) {$b$};
\draw(a22) +(30:1cm) node(c22) {$c$};
\draw[-latex,dashed] (a22) -- (b22);
\draw[-latex] (a22) -- (c22);
\draw[-latex] (b22) -- (c22);
\draw(a22) +(60:5.5mm) node {?};
\draw(a22) +(135:6.5mm) node {$7)$};

\draw(4.8,6) node(a32) {$a$};
\draw(a32) +(90:1cm) node(b32) {$b$};
\draw(a32) +(30:1cm) node(c32) {$c$};
\draw[-latex,dashed] (a32) -- (b32);
\draw[-latex] (c32) -- (a32);
\draw[-latex] (b32) -- (c32);
\draw(a32) +(60:5.5mm) node {$\times$};
\draw(a32) +(135:6.5mm) node {$8)$};

\draw(7.2,6) node(a42) {$a$};
\draw(a42) +(90:1cm) node(b42) {$b$};
\draw(a42) +(30:1cm) node(c42) {$c$};
\draw[-latex,dashed] (a42) -- (b42);
\draw[-latex,dashed] (a42) -- (c42);
\draw[-latex] (b42) -- (c42);
\draw(a42) +(60:5.5mm) node {?};
\draw(a42) +(135:6.5mm) node {$9)$};

\draw(9.6,6) node(a52) {$a$};
\draw(a52) +(90:1cm) node(b52) {$b$};
\draw(a52) +(30:1cm) node(c52) {$c$};
\draw[-latex,dashed] (a52) -- (b52);
\draw[-latex,dashed] (c52) -- (a52);
\draw[-latex] (b52) -- (c52);
\draw(a52) +(60:5.5mm) node {$\times$};
\draw(a52) +(135:6.5mm) node {$10)$};

\draw(0,4) node(a13) {$a$};
\draw(a13) +(90:1cm) node(b13) {$b$};
\draw(a13) +(30:1cm) node(c13) {$c$};
\draw[-latex,dashed] (a13) -- (b13);
\draw[-latex] (c13) -- (b13);
\draw(a13) +(60:5.5mm) node {?};
\draw(a13) +(135:6.5mm) node {$11)$};

\draw(2.4,4) node(a23) {$a$};
\draw(a23) +(90:1cm) node(b23) {$b$};
\draw(a23) +(30:1cm) node(c23) {$c$};
\draw[-latex,dashed] (a23) -- (b23);
\draw[-latex] (a23) -- (c23);
\draw[-latex] (c23) -- (b23);
\draw(a23) +(60:5.5mm) node {$\subsetneqq$};
\draw(a23) +(135:6.5mm) node {$12)$};

\draw(4.8,4) node(a33) {$a$};
\draw(a33) +(90:1cm) node(b33) {$b$};
\draw(a33) +(30:1cm) node(c33) {$c$};
\draw[-latex,dashed] (a33) -- (b33);
\draw[-latex] (c33) -- (a33);
\draw[-latex] (c33) -- (b33);
\draw(a33) +(60:5.5mm) node {$\subsetneqq$};
\draw(a33) +(135:6.5mm) node {$13)$};

\draw(7.2,4) node(a43) {$a$};
\draw(a43) +(90:1cm) node(b43) {$b$};
\draw(a43) +(30:1cm) node(c43) {$c$};
\draw[-latex,dashed] (a43) -- (b43);
\draw[-latex,dashed] (a43) -- (c43);
\draw[-latex] (c43) -- (b43);
\draw(a43) +(60:5.5mm) node {?};
\draw(a43) +(135:6.5mm) node {$14)$};

\draw(9.6,4) node(a53) {$a$};
\draw(a53) +(90:1cm) node(b53) {$b$};
\draw(a53) +(30:1cm) node(c53) {$c$};
\draw[-latex,dashed] (a53) -- (b53);
\draw[-latex,dashed] (c53) -- (a53);
\draw[-latex] (c53) -- (b53);
\draw(a53) +(60:5.5mm) node {?};
\draw(a53) +(135:6.5mm) node {$15)$};

\draw(0,2) node(a14) {$a$};
\draw(a14) +(90:1cm) node(b14) {$b$};
\draw(a14) +(30:1cm) node(c14) {$c$};
\draw[-latex,dashed] (a14) -- (b14);
\draw[-latex,dashed] (b14) -- (c14);
\draw(a14) +(60:5.5mm) node {?};
\draw(a14) +(135:6.5mm) node {$16)$};

\draw(2.4,2) node(a24) {$a$};
\draw(a24) +(90:1cm) node(b24) {$b$};
\draw(a24) +(30:1cm) node(c24) {$c$};
\draw[-latex,dashed] (a24) -- (b24);
\draw[-latex] (a24) -- (c24);
\draw[-latex,dashed] (b24) -- (c24);
\draw(a24) +(60:5.5mm) node {$\interleave$};
\draw(a24) +(135:6.5mm) node {$17)$};

\draw(4.8,2) node(a34) {$a$};
\draw(a34) +(90:1cm) node(b34) {$b$};
\draw(a34) +(30:1cm) node(c34) {$c$};
\draw[-latex,dashed] (a34) -- (b34);
\draw[-latex] (c34) -- (a34);
\draw[-latex,dashed] (b34) -- (c34);
\draw(a34) +(60:5.5mm) node {$\times$};
\draw(a34) +(135:6.5mm) node {$18)$};

\draw(7.2,2) node(a44) {$a$};
\draw(a44) +(90:1cm) node(b44) {$b$};
\draw(a44) +(30:1cm) node(c44) {$c$};
\draw[-latex,dashed] (a44) -- (b44);
\draw[-latex,dashed] (a44) -- (c44);
\draw[-latex,dashed] (b44) -- (c44);
\draw(a44) +(60:5.5mm) node {?};
\draw(a44) +(135:6.5mm) node {$19)$};

\draw(9.6,2) node(a54) {$a$};
\draw(a54) +(90:1cm) node(b54) {$b$};
\draw(a54) +(30:1cm) node(c54) {$c$};
\draw[-latex,dashed] (a54) -- (b54);
\draw[-latex,dashed] (c54) -- (a54);
\draw[-latex,dashed] (b54) -- (c54);
\draw(a54) +(60:5.5mm) node {$\times$};
\draw(a54) +(135:6.5mm) node {$20)$};

\draw(0,0) node(a15) {$a$};
\draw(a15) +(90:1cm) node(b15) {$b$};
\draw(a15) +(30:1cm) node(c15) {$c$};
\draw[-latex,dashed] (a15) -- (b15);
\draw[-latex,dashed] (c15) -- (b15);
\draw(a15) +(60:5.5mm) node {?};
\draw(a15) +(135:6.5mm) node {$21)$};

\draw(2.4,0) node(a25) {$a$};
\draw(a25) +(90:1cm) node(b25) {$b$};
\draw(a25) +(30:1cm) node(c25) {$c$};
\draw[-latex,dashed] (a25) -- (b25);
\draw[-latex] (a25) -- (c25);
\draw[-latex,dashed] (c25) -- (b25);
\draw(a25) +(60:5.5mm) node {$\times$};
\draw(a25) +(135:6.5mm) node {$22)$};

\draw(4.8,0) node(a35) {$a$};
\draw(a35) +(90:1cm) node(b35) {$b$};
\draw(a35) +(30:1cm) node(c35) {$c$};
\draw[-latex,dashed] (a35) -- (b35);
\draw[-latex] (c35) -- (a35);
\draw[-latex,dashed] (c35) -- (b35);
\draw(a35) +(60:5.5mm) node {$\times$};
\draw(a35) +(135:6.5mm) node {$23)$};

\draw(7.2,0) node(a45) {$a$};
\draw(a45) +(90:1cm) node(b45) {$b$};
\draw(a45) +(30:1cm) node(c45) {$c$};
\draw[-latex,dashed] (a45) -- (b45);
\draw[-latex,dashed] (a45) -- (c45);
\draw[-latex,dashed] (c45) -- (b45);
\draw(a45) +(60:5.5mm) node {?};
\draw(a45) +(135:6.5mm) node {$24)$};

\draw(9.6,0) node(a55) {$a$};
\draw(a55) +(90:1cm) node(b55) {$b$};
\draw(a55) +(30:1cm) node(c55) {$c$};
\draw[-latex,dashed] (a55) -- (b55);
\draw[-latex,dashed] (c55) -- (a55);
\draw[-latex,dashed] (c55) -- (b55);
\draw(a55) +(60:5.5mm) node {?};
\draw(a55) +(135:6.5mm) node {$25)$};
\end{tikzpicture}

\caption{Relations between labels $a$, $b$, $c$ with $a\protect\disjorincl b$ for \WPI synthesis.
Some cases are contradictory (shown by a $\times$ sign), while for others
the relation between $a$ and $b$ can be strengthened immediately (to
disjointness $\interleave$ or proper inclusion $\subsetneqq$). If the
edge $a\protect\disjorincl b$ cannot be resolved, there is a question mark
\label{f.25}}
\end{figure}
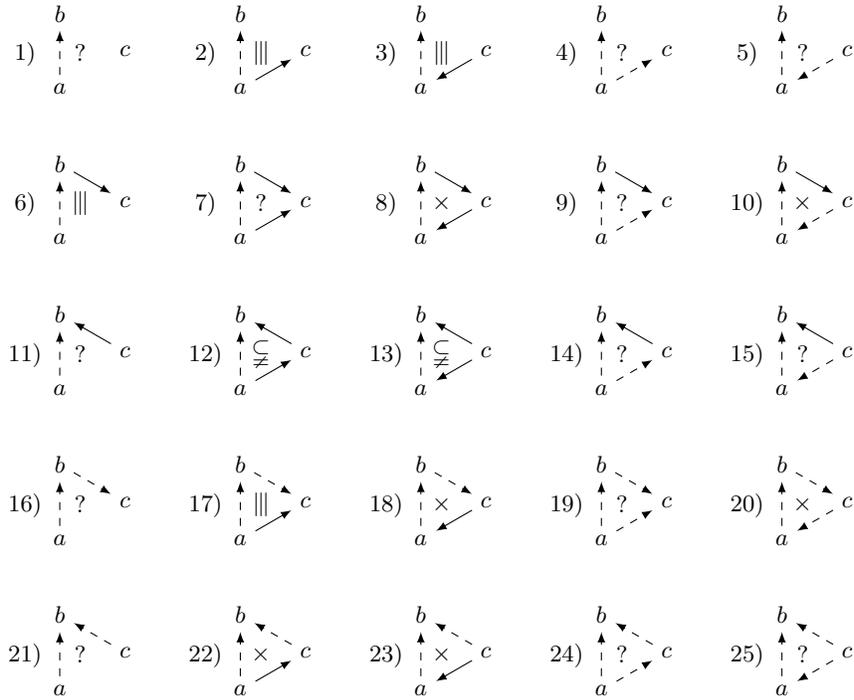

\drop{
\begin{figure}[t]
\centering
\begin{tikzpicture}
\draw(0,4) node(a1) {$a$};
\draw(a1) +(90:1cm) node(b1) {$b$};
\draw(a1) +(30:1cm) node(c1) {$c$};
\draw[-latex,dashed] (a1) -- (b1);
\draw[-latex,dashed] (c1) -- (a1);
\draw[-latex,dashed] (b1) -- (c1);

\draw(0,2) node(a2) {$a$};
\draw(a2) +(90:1cm) node(b2) {$b$};
\draw(a2) +(30:1cm) node(c2) {$c$};
\draw[-latex,dashed] (a2) -- (b2);
\draw[-latex] (c2) -- (a2);
\draw[-latex,dashed] (b2) -- (c2);

\draw(0,0) node(a5) {$a$};
\draw(a5) +(90:1cm) node(b5) {$b$};
\draw(a5) +(30:1cm) node(c5) {$c$};
\draw[-latex,dashed] (a5) -- (b5);
\draw[-latex] (c5) -- (a5);
\draw[-latex,dashed] (c5) -- (b5);

\draw(2,4) node(a3) {$a$};
\draw(a3) +(90:1cm) node(b3) {$b$};
\draw(a3) +(30:1cm) node(c3) {$c$};
\draw[-latex,dashed] (a3) -- (b3);
\draw[-latex,dashed] (c3) -- (a3);
\draw[-latex] (b3) -- (c3);

\draw(2,2) node(a4) {$a$};
\draw(a4) +(90:1cm) node(b4) {$b$};
\draw(a4) +(30:1cm) node(c4) {$c$};
\draw[-latex,dashed] (a4) -- (b4);
\draw[-latex] (c4) -- (a4);
\draw[-latex] (b4) -- (c4);

\draw(2,0) node(a6) {$a$};
\draw(a6) +(90:1cm) node(b6) {$b$};
\draw(a6) +(30:1cm) node(c6) {$c$};
\draw[-latex,dashed] (a6) -- (b6);
\draw[-latex] (a6) -- (c6);
\draw[-latex,dashed] (c6) -- (b6);

\draw[loosely dotted] (3.5,-0.5) -- (3.5,5.5);

\draw(4,0) node(a7) {$a$};
\draw(a7) +(90:1cm) node(b7) {$b$};
\draw(a7) +(30:1cm) node[label=right:$\then$](c7) {$c$};
\draw[-latex,dashed] (a7) -- (b7);
\draw[-latex] (a7) -- (c7);
\draw[-latex] (c7) -- (b7);
\draw(5.7,0) node(a8) {$a$};
\draw(a8) +(90:1cm) node(b8) {$b$};
\draw(a8) +(30:1cm) node(c8) {$c$};
\draw[-latex] (a8) -- (b8);
\draw[-latex] (a8) -- (c8);
\draw[-latex] (c8) -- (b8);

\draw(7.7,0) node(a9) {$a$};
\draw(a9) +(90:1cm) node(b9) {$b$};
\draw(a9) +(30:1cm) node[label=right:$\then$](c9) {$c$};
\draw[-latex,dashed] (a9) -- (b9);
\draw[-latex] (c9) -- (a9);
\draw[-latex] (c9) -- (b9);
\draw(9.4,0) node(a10) {$a$};
\draw(a10) +(90:1cm) node(b10) {$b$};
\draw(a10) +(30:1cm) node(c10) {$c$};
\draw[-latex] (a10) -- (b10);
\draw[-latex] (c10) -- (a10);
\draw[-latex] (c10) -- (b10);

\draw[loosely dotted] (3.5,1.5) -- (10.9,1.5);

\draw(4,2) node(a11) {$a$};
\draw(a11) +(90:1cm) node(b11) {$b$};
\draw(a11) +(30:1cm) node[label=right:$\then$](c11) {$c$};
\draw[-latex,dashed] (a11) -- (b11);
\draw[-latex] (c11) -- (a11);
\draw(5.7,2) node(a12) {$a$};
\draw(a12) +(90:1cm) node(b12) {$b$};
\draw(a12) +(30:1cm) node(c12) {$c$};
\draw[-latex] (c12) -- (a12);

\draw(7.7,2) node(a13) {$a$};
\draw(a13) +(90:1cm) node(b13) {$b$};
\draw(a13) +(30:1cm) node[label=right:$\then$](c13) {$c$};
\draw[-latex,dashed] (a13) -- (b13);
\draw[-latex] (b13) -- (c13);
\draw(9.4,2) node(a14) {$a$};
\draw(a14) +(90:1cm) node(b14) {$b$};
\draw(a14) +(30:1cm) node(c14) {$c$};
\draw[-latex] (b14) -- (c14);

\draw(4,4) node(a15) {$a$};
\draw(a15) +(90:1cm) node(b15) {$b$};
\draw(a15) +(30:1cm) node[label=right:$\then$](c15) {$c$};
\draw[-latex,dashed] (a15) -- (b15);
\draw[-latex] (a15) -- (c15);
\draw(5.7,4) node(a16) {$a$};
\draw(a16) +(90:1cm) node(b16) {$b$};
\draw(a16) +(30:1cm) node(c16) {$c$};
\draw[-latex] (a16) -- (c16);

\draw(7.7,4) node(a17) {$a$};
\draw(a17) +(90:1cm) node(b17) {$b$};
\draw(a17) +(30:1cm) node[label=right:$\then$](c17) {$c$};
\draw[-latex,dashed] (a17) -- (b17);
\draw[-latex] (a17) -- (c17);
\draw[-latex,dashed] (b17) -- (c17);
\draw(9.4,4) node(a18) {$a$};
\draw(a18) +(90:1cm) node(b18) {$b$};
\draw(a18) +(30:1cm) node(c18) {$c$};
\draw[-latex] (a18) -- (c18);
\draw[-latex,dashed] (b18) -- (c18);

\end{tikzpicture}
\caption{Relations between labels $a$, $b$, $c$ with pairwise non-identical presets, 
where the preset of $a$ is disjoint to or contained in that of $b$,
shown by a dashed edge. Solid edges mean containment, non-existing ones disjointness.
Label $c$ can be related in 25 ways to $a$ and $b$, six of which (on the left)
are contradictory, either because \IPS{2} demands acyclicity or because $\neg\DPS$
does not fit with the self-loops for $a$ and $c$ (the lower two pictures). In six more cases,
the exact relation of $a$ and $b$ can be concluded, either disjointness (upper right)
or containment (lower right)
\label{f.rel}}
\end{figure}
} 

Figure~\ref{f.25} also provides strengthenings of $\disjorincl$ to $\included$ via 12) and 13). 
The strengthened edge can be part of another triangle, as shown in Fig.~\ref{f.indirect}.
In all six situations, the edge $a\disjorincl b$ is strengthened to $a\included b$ via 12) or
13). The triangle consisting of $a$, $b$, and $d$ is now either that of 22) or 23), but we
cannot draw a conclusion directly. For 22) and 23) it is essential that $a$ and $b$ (named
$a$ and $c$ in Fig.~\ref{f.25}) can deactivate each other, but here that is not the case.
We can still draw the same conclusion in an indirect way. Take the middle picture in the
upper row of Fig.~\ref{f.indirect} as an example: We have triangles $acd$ and $bcd$ to
which 22) or 23) can be applied, showing the inconsistency we originally wanted to derive
from the triangle $abd$. In this way, if an edge $\included$ stems from a strengthening
disallowing the application of 22) or 23), we fall back to the triangle responsible for
the strengthened edge. This can even be done over several steps. Note that there may be
different conclusions though: In the pictures on the left and right hand sides of Fig.~\ref{f.indirect}
we get disjointness and proper inclusion as a result, respectively.

\begin{figure}[t]
\centering
\begin{tikzpicture}
\draw(0,2) node(a1) {$a$};
\draw(a1) +(90:1cm) node(b1) {$b$};
\draw(a1) +(30:1cm) node(c1) {$c$};
\draw(a1) +(150:1cm) node(d1) {$d$};
\draw[-latex,dashed] (a1) -- (b1);
\draw[-latex] (a1) -- (c1);
\draw[-latex] (c1) -- (b1);
\draw[-latex,dashed] (b1) -- (d1);
\draw[-latex,dashed] (a1) -- (d1);
\draw(a1) +(60:5.5mm) node {$\subsetneqq$};
\draw(a1) +(120:5.5mm) node {$\interleave$};
\draw(d1) +(-0.5,0.5) node {$12)+2)+3)$};

\draw(4,2) node(a3) {$a$};
\draw(a3) +(90:1cm) node(b3) {$b$};
\draw(a3) +(30:1cm) node(c3) {$c$};
\draw(a3) +(150:1cm) node(d3) {$d$};
\draw[-latex,dashed] (a3) -- (b3);
\draw[-latex] (a3) -- (c3);
\draw[-latex] (c3) -- (b3);
\draw[-latex,dashed] (b3) -- (d3);
\draw[-latex,dashed] (a3) -- (d3);
\draw[-latex,dashed] (c3) edge[bend left=90] (d3);
\draw(a3) +(60:5.5mm) node {$\subsetneqq$};
\draw(a3) +(120:5.5mm) node {$\times$};
\draw(d3) +(-0.5,0.5) node {$12)+22)+23)$};

\draw(8,2) node(a5) {$a$};
\draw(a5) +(90:1cm) node(b5) {$b$};
\draw(a5) +(30:1cm) node(c5) {$c$};
\draw(a5) +(150:1cm) node(d5) {$d$};
\draw[-latex,dashed] (a5) -- (b5);
\draw[-latex] (a5) -- (c5);
\draw[-latex] (c5) -- (b5);
\draw[-latex,dashed] (b5) -- (d5);
\draw[-latex,dashed] (a5) -- (d5);
\draw[-latex] (c5) edge[bend left=90] (d5);
\draw(a5) +(60:5.5mm) node {$\subsetneqq$};
\draw(a5) +(120:5.5mm) node {$\subsetneqq$};
\draw(d5) +(-0.5,0.5) node {$12)+12)+13)$};

\draw(0,0) node(a2) {$a$};
\draw(a2) +(90:1cm) node(b2) {$b$};
\draw(a2) +(30:1cm) node(c2) {$c$};
\draw(a2) +(150:1cm) node(d2) {$d$};
\draw[-latex,dashed] (a2) -- (b2);
\draw[-latex] (c2) -- (a2);
\draw[-latex] (c2) -- (b2);
\draw[-latex,dashed] (b2) -- (d2);
\draw[-latex,dashed] (a2) -- (d2);
\draw(a2) +(60:5.5mm) node {$\subsetneqq$};
\draw(a2) +(120:5.5mm) node {$\interleave$};
\draw(d2) +(-0.5,0.5) node {$13)+3)+3)$};

\draw(4,0) node(a4) {$a$};
\draw(a4) +(90:1cm) node(b4) {$b$};
\draw(a4) +(30:1cm) node(c4) {$c$};
\draw(a4) +(150:1cm) node(d4) {$d$};
\draw[-latex,dashed] (a4) -- (b4);
\draw[-latex] (c4) -- (a4);
\draw[-latex] (c4) -- (b4);
\draw[-latex,dashed] (b4) -- (d4);
\draw[-latex,dashed] (a4) -- (d4);
\draw[-latex,dashed] (c4) edge[bend left=90] (d4);
\draw(a4) +(60:5.5mm) node {$\subsetneqq$};
\draw(a4) +(120:5.5mm) node {$\times$};
\draw(d4) +(-0.5,0.5) node {$13)+23)+23)$};

\draw(8,0) node(a6) {$a$};
\draw(a6) +(90:1cm) node(b6) {$b$};
\draw(a6) +(30:1cm) node(c6) {$c$};
\draw(a6) +(150:1cm) node(d6) {$d$};
\draw[-latex,dashed] (a6) -- (b6);
\draw[-latex] (c6) -- (a6);
\draw[-latex] (c6) -- (b6);
\draw[-latex,dashed] (b6) -- (d6);
\draw[-latex,dashed] (a6) -- (d6);
\draw[-latex] (c6) edge[bend left=90] (d6);
\draw(a6) +(60:5.5mm) node {$\subsetneqq$};
\draw(a6) +(120:5.5mm) node {$\subsetneqq$};
\draw(d6) +(-0.5,0.5) node {$13)+13)+13)$};
\end{tikzpicture}
\caption{How to resolve 22)/23) of Fig.~\ref{f.25} when the edge $a\protect\included b$
in the triangle $a$/$b$/$d$ is created from strengthening $a\protect\disjorincl b$ via $c$ and 12) or 13).
All possible relations between $c$ and $d$ are shown.
The $\subsetneqq$ sign in the right triangle denotes the strengthening of the $a\protect\disjorincl b$ edge,
the sign in the left triangle belongs to both edges leading to $d$, $a\protect\disjorincl d$
and $b\protect\disjorincl d$. The numbers at each picture show which parts of Fig.~\ref{f.25} are applied
to the three triangles containing $c$
\label{f.indirect}}
\end{figure}

Note how the middle column of Fig.~\ref{f.25} is now fully resolved: If there is no contradiction
prohibiting synthesis, we know exactly whether the edge $a\disjorincl b$ stands for disjointness
or for proper inclusion. An unresolved situation $c\included a\disjorincl b$ does not occur anymore.
As a consequence, a self-loop $a$ with $a\disjorincl b$ can only be deactivated in a synthesised
\WPI net by a label/transition whose preset encompasses that of $a$.
Since 7) is the only unresolved situation in the second column,
a proper inclusion edge starting at $a$ implies a proper inclusion edge at $b$ to the
same target label.
Unfortunately, we find no more simple conclusions like this one as every other row and column of the figure
still contains more than one unresolved situation.

The above conclusions reduce the number of $\disjorincl$ edges in our relations, but
some of them may remain. For all such edges, the possibilities, disjointness and
proper inclusion, have to be tested all in conjunction when constructing inequality systems 
for synthesis.
Since the number of these edges is not limited in general, synthesis to \WPI nets 
may need exponential time with respect to them.
We might argue that we expect relatively few labels compared to states ($O(|\Sigma|)\subseteq O(\log|S|)$)
so that the synthesis keeps its polynomial time complexity, or that we have only
one `idle' label forming a self-loop in the \lts, giving us just a constant factor
as overhead. We can reduce the problem by the shown strenghtening of edges, which
can be computed in $O(|S|\cdot|\Sigma|^2+|\Sigma|^3)$ compared to roughly $O(|S|^6)$ for solving a
single inequality system in the rational numbers~\cite{karmarkar}. But in the end,
we still have a potentially exponential problem. We may guess for each edge
$\disjorincl$ if it means disjointness or proper inclusion, bringing us into
the class \NP.

\CON{c.1}{Self-loops might make \WPI synthesis exponential}\\
If an \lts has no self-loops or just one, synthesis to the target class \WPI
can be done in polynomial time. With an arbitrary number of labels forming self-loops,
we rather expect the problem to be \NP-complete,
but we also expect `realistic' \lts to fit into the first category.
\ENDCON{c.1}

\section{Self-loops and the class \BRAC}\label{sect.5}

When targetting our second class of nets, \BRAC, tackling self-loops is considerably
easier. Any asymmetric choice block of transitions consists of two subblocks, where
one has a single, common place as preset, and the other has this place and one other
as a common preset. Two asymmetric choice blocks do not have common transitions.
Therefore, the situations shown in Fig.~\ref{f.ntbrac} are impossible when targetting
at \BRAC nets.%
\footnote{
For \ACC nets only the third situation in Fig.~\ref{f.ntbrac} is impossible as all places
in ${}^\bullet a\cup{}^\bullet b$ have the common postset transition $c$ and must thus have
comparable postsets. The fifth situation can easily occur when $a$, $b$, and $c$ share a
common preset place but $b$ and $c$ also have unshared preset places.
}

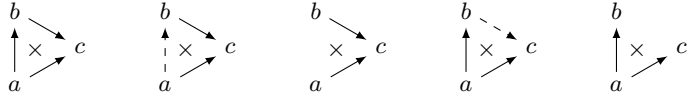
\begin{figure}[t]
\centering
\begin{tikzpicture}
\draw(0,0) node(a1) {$a$};
\draw(a1) +(90:1cm) node(b1) {$b$};
\draw(a1) +(30:1cm) node(c1) {$c$};
\draw[-latex] (a1) -- (b1);
\draw[-latex] (a1) -- (c1);
\draw[-latex] (b1) -- (c1);
\draw(a1) +(60:5.5mm) node {$\times$};

\draw(2,0) node(a2) {$a$};
\draw(a2) +(90:1cm) node(b2) {$b$};
\draw(a2) +(30:1cm) node(c2) {$c$};
\draw[-latex,dashed] (a2) -- (b2);
\draw[-latex] (a2) -- (c2);
\draw[-latex] (b2) -- (c2);
\draw(a2) +(60:5.5mm) node {$\times$};

\draw(4,0) node(a3) {$a$};
\draw(a3) +(90:1cm) node(b3) {$b$};
\draw(a3) +(30:1cm) node(c3) {$c$};
\draw[-latex] (a3) -- (c3);
\draw[-latex] (b3) -- (c3);
\draw(a3) +(60:5.5mm) node {$\times$};

\draw(6,0) node(a4) {$a$};
\draw(a4) +(90:1cm) node(b4) {$b$};
\draw(a4) +(30:1cm) node(c4) {$c$};
\draw[-latex] (a4) -- (b4);
\draw[-latex] (a4) -- (c4);
\draw[-latex,dashed] (b4) -- (c4);
\draw(a4) +(60:5.5mm) node {$\times$};

\draw(8,0) node(a5) {$a$};
\draw(a5) +(90:1cm) node(b5) {$b$};
\draw(a5) +(30:1cm) node(c5) {$c$};
\draw[-latex] (a5) -- (b5);
\draw[-latex] (a5) -- (c5);
\draw(a5) +(60:5.5mm) node {$\times$};
\end{tikzpicture}
\caption{For \BRAC, proper inclusions with more than two labels involved are possible only if two labels are equivalent.
It does not matter whether the proper inclusion arrows are consecutive or begin or end at the same label
\label{f.ntbrac}}
\end{figure}

When looking at Fig.~\ref{f.25}, this means that in 7), 9), 11), 14), and 15) the edge $a\disjorincl b$ needs
to represent disjointness and can be strengthened accordingly. In the remaining unresolved cases
1), 4), 5), 16), 19), 21), 24), and 25) all labels only have edges $\disjorincl$ or show disjointness,
proper inclusion edges do not occur anymore. Thus, the transitions in a synthesised \BRAC net corresponding
to these labels form a block with presets completely disjoint to all other transitions
(aside from equivalent labels, which we have ignored).

For two edges $a\disjorincl b\disjorincl c$ we do not see immediately which of the two edges enforces
proper inclusion of presets (if any), but clearly not both of them can have this meaning.
A closer look shows:

\LEM{l.firstdoiisdisj}{The front edge $\disjorincl$ means disjointness}
Let $TS=(S,\Sigma,\to,s_0)$ be a finite, reachable, deterministic \lts with label relations
strengthened according to Section~\ref{sect.4}.
Let there be three labels $a,b,c\in\Sigma$ with $a\disjorincl b\disjorincl c$. 
Then, if $TS$ is synthesisable to \BRAC, we can always find a \BRAC net $N=(P,T,W,M_0)$ solving $TS$ with 
${}^\bullet a\cap{}^\bullet b=\emptyset$.
\ENDLEM
\BEW
Note first, that by case~6 $a$ and $b$ are both self-loops in $TS$. There may be labels
equivalent to $a$ or $b$, but by Lemma~\ref{l.equiv} these also all are self-loops in $TS$.
Assume now that $a\disjorincl b$ stands for proper inclusion, otherwise we are done.
Since $a$ and $b$ and their equivalents are then related to all other non-equivalent labels by disjointness,
no other transition $t$ in a \BRAC net solving $TS$ can take tokens from a place $p$
in the preset of $a$ ($W(p,a)>0\then a\equivalent t\vee b\equivalent t\vee W(p,t)=0$). Assume a region $r=(R,B,F)$
solving an ESSP $(s,a)$ or an SSP $(s,s')$ with $s,s'\in S$. Then $B(b)=F(b)$, and with
$B'(b')=F'(b')=0$ for all $b'\in\Sigma$ with $b\equivalent b'$ and $B'(t)=B(t)\wedge F'(t)=F(t)$ for all
other labels $t$, the region
$r'=(R,B',F')$ is also a solution for the inequality system for the same ESSP or SSP.
Thus, by making a disjoint copy of the preset of $a$ and disconnecting $a$ and $b$ (and their equivalent labels) from
each other's new preset, we obtain a solution for $TS$ with ${}^\bullet a\cap{}^\bullet b=\emptyset$.
Clearly, the \BRAC conditions still hold, as there are no other (non-equivalent) transitions involved
that take tokens from the preset of $a$ (old or new). 
\ENDBEW{l.firstdoiisdisj}

We can now, in every chain $a\disjorincl b\disjorincl c$, replace the first edge by disjointness
simultaneously. What remains are two sets $A$ and $B$ of labels, such that inside the sets
$\forall a,a'\in A\colon a\disjoint a'\vee a\equivalent a'$ and
$\forall b,b'\in B\colon b\disjoint b'\vee b\equivalent b'$ hold
while between the sets we have
$\forall a\in A, b\in B\colon a\disjoint b\vee a\disjorincl b$.

\tikzset{rloop/.style={to path={.. controls +(325:1) and +(35:1) .. (\tikztotarget) \tikztonodes}}}
\tikzset{lloop/.style={to path={.. controls +(145:1) and +(215:1) .. (\tikztotarget) \tikztonodes}}}
\tikzset{dloop/.style={to path={.. controls +(235:1) and +(305:1) .. (\tikztotarget) \tikztonodes}}}
\tikzset{drloop/.style={to path={.. controls +(280:1) and +(350:1) .. (\tikztotarget) \tikztonodes}}}
\tikzset{tloop/.style={to path={.. controls +(55:1) and +(125:1) .. (\tikztotarget) \tikztonodes}}}
\tikzset{trloop/.style={to path={.. controls +(0:1) and +(70:1) .. (\tikztotarget) \tikztonodes}}}
\tikzset{tlloop/.style={to path={.. controls +(160:1) and +(90:1) .. (\tikztotarget) \tikztonodes}}}
\begin{figure}[t]
\centering
\begin{tikzpicture}[LTS]
\node[state,label=above:$s_0$] (s0) {};
\path[edge] (s0) ++(-0.5,0) edge (s0);
\node[state,right of=s0,label=right:$s_1$] (s1) {};
\draw[edge] (s0) edge node[auto,xshift=-1mm]{$a$} (s1);
\draw[edge] (s1) edge[tlloop] node[auto,swap]{$c$} (s1);
\node[state,above right of=s1,label=below:$s_2$] (s2) {};
\draw[edge] (s1) edge node[auto,xshift=1.5mm]{$a$} (s2);
\node[state,below right of=s1,label=above:$s_3$] (s3) {};
\draw[edge] (s1) edge node[auto,swap,xshift=0.5mm,yshift=0.5mm]{$b$} (s3);
\node[state,below right of=s2,label=left:$s_4$] (s4) {};
\draw[edge] (s2) edge node[auto,xshift=-1.5mm]{$b$} (s4);
\draw[edge] (s2) edge[tloop] node[auto]{$c$} (s2);
\draw[edge] (s3) edge node[auto,swap,xshift=-0.5mm]{$a$} (s4);
\draw[edge] (s3) edge[drloop] node[auto]{$c$} (s3);
\node[state,below right of=s4,label=below:$s_5$] (s5) {};
\draw[edge] (s4) edge node[auto,xshift=-0.5mm,yshift=-1mm]{$b$} (s5);
\draw[edge] (s4) edge[trloop] node[auto]{$c$} (s4);
\draw[edge] (s5) edge[rloop] node[auto]{$c$} (s5);
\node[state,below left of=s1,label=above:$s_6$] (s6) {};
\draw[edge] (s4) .. controls +(0.5,-2.3) and +(0,-2.3) .. node[auto]{$d$} (s6);
\draw[edge] (s6) edge node[auto,swap,yshift=0.5mm]{$e$} (s1);
\node[state,below right of=s6,label=left:$s_7$] (s7) {};
\draw[edge] (s6) edge node[auto,yshift=-0.5mm]{$b$} (s7);
\draw[edge] (s7) edge node[auto,yshift=-0.5mm]{$e$} (s3);
\draw[edge] (s6) edge[lloop] node[auto]{$c$} (s6);
\draw[edge] (s7) edge[drloop] node[auto]{$c$} (s7);
\end{tikzpicture}\hspace*{1cm}
\begin{tikzpicture}
\draw(0,3) node[place,label=left:$p_1$,tokens=2] (p1) {};
\draw(0,2) node[place,label=left:$p_2$] (p2) {};
\draw(0,0) node[place,label=left:$p_3$] (p3) {};
\draw(0,1) node[transition] (b) {$b$};
\draw(1.5,2.25) node[transition] (a) {$a$};
\draw(1.5,0.75) node[transition] (d) {$d$};
\draw(3,2.25) node[place,label=below:$p_4$] (p4) {};
\draw(3,0.75) node[place,label=above:$p_5$] (p5) {};
\draw(4.5,2.25) node[transition] (c) {$c$};
\draw(5.5,1.5) node[transition] (e) {$e$};
\draw(p1) edge[-latex,thick] (a);
\draw(a) edge[-latex,thick] (p2);
\draw(p2) edge[-latex,thick] (b);
\draw(b) edge[-latex,thick] (p3);
\draw(p2) edge[-latex,thick,bend left=20] (d);
\draw(d) edge[-latex,thick,bend left=20] (p2);
\draw(p3) edge[-latex,thick] (d);
\draw(d) edge[-latex,thick] (p5);
\draw(a) edge[-latex,thick] (p4);
\draw(p4) edge[-latex,thick,bend left=20] (c);
\draw(c) edge[-latex,thick] (p4);
\draw(p4) edge[-latex,thick] (e);
\draw(p5) edge[-latex,thick] (e);
\draw(e) edge[-latex,thick,out=90,in=0] (p1);
\path(p3) +(0,-1);
\end{tikzpicture}\vspace*{-5mm}
\caption{An \lts synthesisable to the shown (B)\RAC net. Note that $s_0\step{a}s_1$ activates $c$ and $a$ must
put a token in $c$'s preset. The cycle $s_1\step{abde}s_1$ is token-neutral, so one of its transitions
must remove a token from $c$'s preset}
\label{f.bracnodisj}
\end{figure}
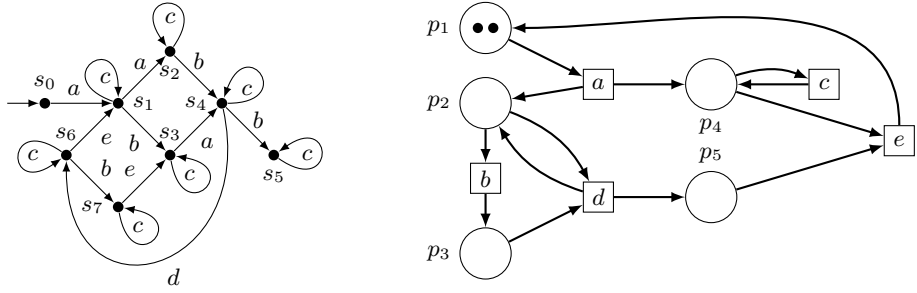

Unfortunately, we cannot choose all the remaining $a\disjorincl b$ edges to mean disjointness.
Figure~\ref{f.bracnodisj} shows an \lts together with a \BRAC (and also \RAC) net solving the \lts.
We find $c\disjorincl b$, $c\disjorincl d$, $c\disjorincl e$, and $b\included d$,
all other pairs are not connected. Now, $b\included d$ forces $c$ to be disjoint from $b$ and $d$,
otherwise we obtain one of the situations of Fig.~\ref{f.ntbrac}. If $c$ and $e$ also have
disjoint presets in a synthesised \BRAC net, we will conclude that the cycle $s_1\step{abde}s_1$
cannot take tokens from $c$'s preset, so it cannot put tokens there either. But $a$ must put a
token there as $s_0\step{a}s_1$ activates $c$. So, ${}^\bullet c\subsetneqq{}^\bullet e$ holds in
every \BRAC net solving this \lts. As we cannot enforce disjointness, we must compute whether
any edge $\disjorincl$ means proper inclusion or if disjointness is possible.

\SATZ{l.esspbrac}{Linear number of inequality systems for an ESSP} 
For a finite, reachable, deterministic \lts $TS=(S,\Sigma,\to,s_0)$ assume label relations
to be constructed according to Def.~\ref{d.grl} and strengthened via Lemma~\ref{l.equiv}, Fig.~\ref{f.25},
Fig.~\ref{f.ntbrac}, and Lemma~\ref{l.firstdoiisdisj}.
When targetting at \BRAC,
for each ESSP $(s,a)$ with $s\in S$ and $a\in\Sigma$ we need to solve at most $|\Sigma|$ inequality systems.
\ENDSATZ
\BEW
If there is no edge $\disjorincl$ or $\included$ adjacent to $a$ in the label relations, 
$a$ is not involved in any asymmetric choice. Each ESSP $(s,a)$ with $s\in S$ can then be solved
separately as $a$ can have as many places in its preset as necessary. If any of these
inequality systems is unsolvable, the \lts is not synthesisable to a \BRAC net.

If an edge $\included$ starts or ends at $a$, $a$ is in an asymmetric choice for certain.
All ESSPs $(s,a)$ for $s\in S$ must be solved by the same inequality system, as shown in
Section~\ref{sect.3}. If the inequality system is not solvable, synthesis fails.

If an edge $b\disjorincl a$ exists with $b\in\Sigma$, $b$ is a self-loop and firing $b$
in a synthesised \BRAC net (if it exists) will not change the net's marking, especially
in the preset of $a$. 
We check whether we are able to construct a preset for $a$ if all edges
$t\disjorincl a$ with $t\in\Sigma$ are interpreted as disjointness, trying to solve one
inequality system per ESSP. If this is not possible,
$TS$ cannot be synthesised at all. Checking whether any edge $\disjorincl$ ending at $a$ must mean proper
inclusion is postponed to the following case.

Assume now $a\disjorincl b$ for some $b\in\Sigma$. We try to interpret all edges $\disjorincl$
starting at $a$ as disjointness and solve each ESSP $(s,a)$ with $s\in S$ separately.
Suppose this is possible, i.e.\ all these ESSPs are solvable.
Since $a$ cannot change any token distribution in a synthesised net, this will not help $b$ in any way.
As $b$ is arbitrary, sharing preset places with $a$ never helps in solving other ESSPs. We may therefore
choose to make a free-choice block of $a$ and its equivalents in a \BRAC net.

If not all ESSPs $(s,a)$ are solvable when interpreting all edges $a\disjorincl$ as disjointness,
one of the edges, $a\disjorincl b$ for some $b\in\Sigma$ must mean proper inclusion,
${}^\bullet a\subsetneqq{}^\bullet b$, but we do not know which edge.
For each possibility, we construct one inequality system for all ESSPs $(s,a)$ and try to solve it.
If the inequality system for $a\disjorincl b$ meaning proper inclusion is not solvable, the edge
must be interpreted as disjointness. We collect all pairs $(a,b)$ where proper inclusion for
the edge $a\disjorincl b$ leads to a solvable inequality system in a set $\Lambda$.

After dealing with all the edges $\disjorincl$ in this way, $\Lambda$ contains a number of pairs $(t,t')$. If, for a given $t$, for two pairs
$(t,t'),(t,t'')\in \Lambda$ holds $t'\equivalent t''$, 
we choose $t'$ as a representative for the equivalence and remove $(t,t'')$ from $\Lambda$.
If for some $t$ there remain pairs $(t,t'),(t,t'')\in\Lambda$ where $t'$ and $t''$ are not
equivalent, we must find out if the proper inclusion should hold towards $t'$ or $t''$
(or possibly another label). This is a bipartite matching problem: Given the labels and
edges occuring in the bipartite graph formed by $\Lambda$, we must select
$|\{a\mid (a,b)\in\Lambda\}|$ edges such that no label (i.e.\ node of the graph) is
adjacent to more than one of them. Due to the graph structure, $|\{a\mid (a,b)\in\Lambda\}|$
is also the maximal number of edges that can ever be selected with this property.
This maximum bipartite matching problem~\cite{mv80} can be solved in polynomial time, in
our case in $O(|\Sigma|^{2.5})$. If the maximum matching has exactly $|\{a\mid (a,b)\in\Lambda\}|$
edges, all our problems are solved, we know which edges $\disjorincl$ mean proper inclusion, 
and we refer to the previously computed solutions for our inequality
system to construct the places for the synthesised \BRAC net. 

The maximal number of inequality systems to solve for a single ESSP $(s,a)$ is then $|\Sigma|$, 
one to test for disjointness and (at most) one for each possible edge $a\disjorincl b$ 
or $b\disjorincl a$ with $b\in\Sigma\backslash\{a\}$.
\ENDBEW{l.esspbrac}

Note that there are still two sources for a possibly non-polynomial runtime, one of which
we cannot avoid:
Integrating SSPs into the inequality systems for ESSPs (as explained in Section~\ref{sect.3}) 
may still lead to an exponential number of inequality systems
and/or solving inequality systems remains \NP-complete for the class \BRAC.

\section{Conclusion}\label{sect.O}

We have shown that synthesis of an \lts to obtain a Petri net from the class \WPI can be done
in polynomial time if the \lts contains only a fixed number of labels forming self-loop edges, 
i.e. cycles of length one. Otherwise, we only have an exponential-time algorithm in \NP. 
The true goal of this paper was the synthesis to nets in some class of asymmetric choice,
though. Due to the nature of the asymmetric choice condition, synthesis to the general class
of asymmetric choice nets (\ACC) cannot even be formulated in the context of separation problems.
For one subclass of \ACC, \BRAC (block-reduced asymmetric choice nets), we
have shown that the synthesis can be done at all, but we have also identified several sources
of a possible non-polynomial runtime. If we allow arc weights greater than one (reducing
problems of solving inequality systems to rational solutions with a polynomial runtime),
we run into combinatorial problems with state separation (SSPs). Self-loops, which posed
the biggest problem for \WPI synthesis, can essentially be dealt with in polynomial time
for the class \BRAC.

\end{document}